\documentclass[sigconf]{acmart}

\usepackage[normalem]{ulem}

\usepackage{amsmath}

\usepackage{amssymb}

\usepackage{physics}
\usepackage{algorithm}
\usepackage{algpseudocode}
\usepackage{graphicx}
\usepackage{environ}
\usepackage{textcomp}
\usepackage{subfig}

\acmConference[]{arXiv}{August}{2021}
\renewcommand\footnotetextcopyrightpermission[1]{}

\newcommand{\sol}{\textsc{QEst}}
\NewEnviron{gather+}[1][1]{%
  \begin{equation}
  \scalebox{#1}{$\begin{gathered}\BODY\end{gathered}$}
  \end{equation}
}

\title{Robust and Resource-Efficient Quantum Circuit Approximation}

\date{}

\author{Tirthak Patel}
\affiliation{
  \institution{Northeastern University}\city{Boston}
  \state{Massachusetts}
  \country{USA}
  }
  
\author{Ed Younis}
\affiliation{
  \institution{Lawrence Berkeley National Laboratory}\city{Berkeley}
  \state{California}
  \country{USA}
  }
  
\author{Costin Iancu}
\affiliation{
  \institution{Lawrence Berkeley National Laboratory}\city{Berkeley}
  \state{California}
  \country{USA}
  }
  
\author{Wibe de Jong}
\affiliation{
  \institution{Lawrence Berkeley National Laboratory}\city{Berkeley}
  \state{California}
  \country{USA}
  }
  
\author{Devesh Tiwari}
\affiliation{
  \institution{Northeastern University}\city{Boston}
  \state{Massachusetts}
  \country{USA}
  }

\begin{document}

\begin{abstract}

We present \sol{}, a procedure to systematically generate approximations for quantum circuits to reduce their CNOT gate count. Our approach employs circuit partitioning for scalability with procedures to 1) reduce circuit length using approximate synthesis, 2) improve fidelity by running circuits that represent key samples in the approximation space, and 3) reason about approximation upper bound. Our evaluation results indicate that our approach of ``dissimilar'' approximations provides close fidelity to the original circuit. Overall, the results indicate that \sol{} can reduce CNOT gate count by 30-80\% on ideal systems and decrease the impact of noise on existing and near-future quantum systems.


\end{abstract}
\maketitle

\section{Introduction and Motivation}
\label{sec:introduction}

We begin by providing a brief relevant background of quantum computing, followed by motivation for \sol{}, and summarize the approach and contributions of \sol{}.

\subsection{Quantum Computing: A Brief Background}

The unit of information in quantum computing is the qubit. This is a two-level abstraction, whose state is represented by $\ket{\psi} = \alpha\ket{0} + \beta\ket{1}$, which is a superposition of the $\ket{0}$ and $\ket{1}$ states. When this qubit is measured, its superposition collapses and it can be measured in state $\ket{0}$ with probability $\norm{\alpha}^2$ and in state $\ket{1}$ with probability $\norm{\beta}^2$, such that $\norm{\alpha}^2 + \norm{\beta}^2 = 1$. Similarly, an $n$-qubit entangled quantum system can be represented as a superposition of $2^n$ states: $\ket{\psi} = \displaystyle\sum_{k=0}^{k=n-1}\alpha_k\ket{k}$, such that $\displaystyle\sum_{k=0}^{k=n-1}\norm{\alpha_k}^2 = 1$.

Under Deutch's computational model, a quantum program can be represented as a circuit of operations~\cite{deutsch1989quantum}. A quantum system can be put into a desired superposition state using quantum gates or operations. For example, a one-qubit NOT operation can be used to rotate the state of qubit ($\ket{0}$ to $\ket{1}$ or vice versa), while a two-qubit controlled-NOT or CNOT operation can be used to entangle two qubits and apply a NOT gate to the ``target'' qubit if the ``control'' qubit is in the $\ket{1}$ state. When these operations are applied in succession to one another, they form a ``quantum circuit'' that represents and executes the corresponding ``quantum algorithm.'' Note that all quantum algorithms can be represented as a sequence of one-qubit rotation gates and two-qubit CNOT gates.

\subsection{Motivation for \sol{}}

\begin{figure}[t]
    \centering
    \subfloat[][TFIM]{\includegraphics[scale=0.43]{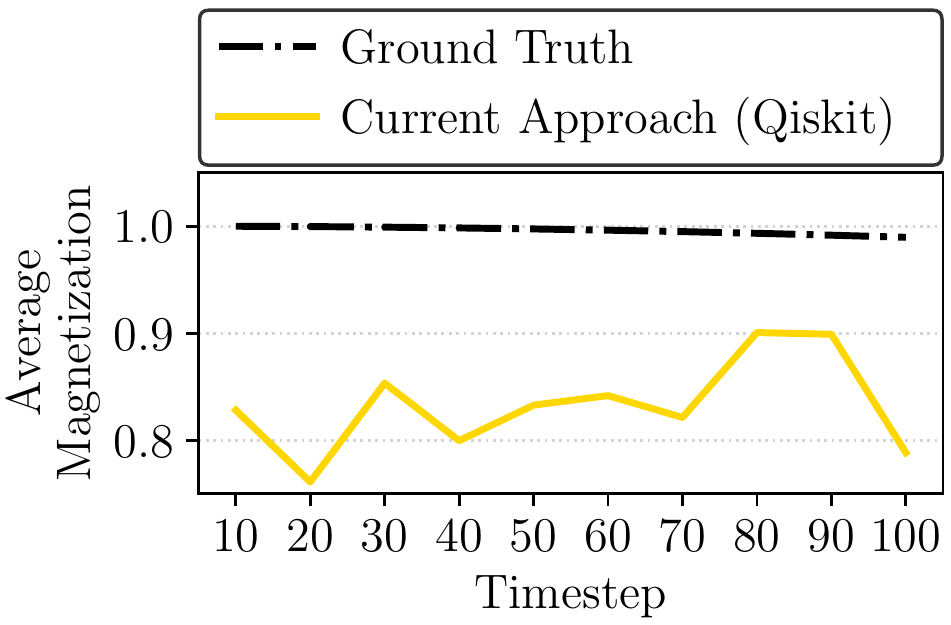}}\hfill
    \subfloat[][Heisenberg]{\includegraphics[scale=0.43]{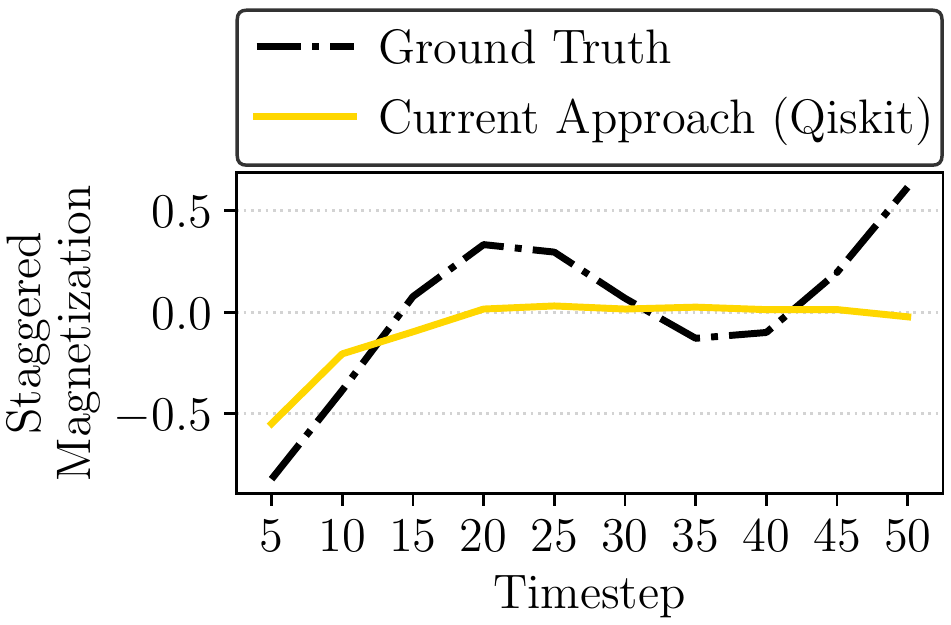}}
    \caption{The output of the TFIM and Heisenberg quantum algorithms on the IBMQ Manila quantum computer using all Qiskit compiler optimizations (current approach) is far from the expected output (ground truth) at different timesteps.}
    \label{fig:tfim_heis_mot}
  \end{figure}

Existing quantum systems and projected near-term future quantum systems suffer from a pack of different types of errors~\cite{li2020qasmbench,preskill2018quantum}. These include state-preparation and measurement (SPAM) errors and qubit state decoherence errors. These near-term intermediate-scale quantum (NISQ) devices also suffer from errors related to applying one-qubit and two-qubit gates to a qubit. These errors get compounded over the course of a quantum circuit, as a circuit runs many quantum operations. The two-qubit (CNOT) operation error rate (1-3\%) tends to be an order of magnitude above one-qubit operation error rate.

This makes it difficult to run long quantum circuits with many CNOT gates on near-term quantum computers as the CNOT gates have a high application error as well as take longer to apply, causing decoherence errors. Previous works have proposed several techniques that focus on circuit compilation and mapping to hardware in a manner that reduces the CNOT gate count of the circuit as well as reduces the impact of the errors. These include reducing gate count by collapsing adjacent gates, deleting gate operations using commutativity and unitary laws, reducing the number of CNOT and SWAP operations (implemented using three CNOT gates) by performing layout-aware mapping of the quantum circuit to the hardware, and performing noise-aware mapping to ensure that more noisy qubits and operations are avoided as much as possible~\cite{zhang2021time,murali2020software,murali2019noise,tannu2019not,shi2019optimized,li2019tackling,wille2019mapping,zulehner2019compiling,smith2019quantum}. Qiskit~\cite{mckay2018qiskit}, IBMQ's python-based quantum compiling package has these state-of-the-art compilation techniques implemented as optimization passes, whereby the optimizations are applied one after another. While these compilation and mapping passes are effective in some cases, they are not able to reduce the gate count to a degree that is able to reduce the noise sufficiently.

As an instance, Fig.~\ref{fig:tfim_heis_mot}(a) and (b) show the output of the TFIM (Transverse Field Ising Model) and Heisenberg quantum algorithms, respectively, in the ideal (ground truth) scenario vs. the case in which they are on run on the real IBMQ Manila quantum computer~\cite{castelvecchi2017ibm}. TFIM outputs the time evolution of the average magnetization (energy) of a four-spin physical system with only the $z$ Hamiltonian interaction component, while Heisenberg outputs the time evolution of a four-spin system with $x, y, z$ Hamiltonian interaction components~\cite{bassman2021arqtic}. Even though the IBMQ Manila computer is a relatively low-error NISQ device and all of the Qiskit compiler optimizations are applied when running the circuit on the computer, the output is far from the expected ideal output. The error is large enough for the output to be not provide meaningful insights. For example, with TFIM, the output does not remain consistent across the timesteps (even though it should according to the ground truth output), not does it have the same magnetization amplitude as the ground truth.

\subsection{Approach and Contributions of \sol{}}

Overall, the compilation and mapping passes employed by the state-of-the-art approaches cannot reduce the output error sufficiently. Therefore in this work, we present \sol{}, a technique that focuses on delivering a large reduction in CNOT gate count to reduce the effect of noise.

\sol{} is built on are a few key observations and opportunities. A quantum circuit can be mathematically represented as a unitary matrix. A unitary matrix can have multiple mathematically close ``approximations''. These approximated unitaries can represent the original circuit's functionality. \sol{} leverages this property to demonstrate that it is possible to use approximations to our advantage to improve the output fidelity of complex quantum programs on erroneous NISQ devices.

While the idea of approximating circuits appears promising, it poses multiple challenges not solved by prior works before it can be realized in practice. First, approximating a complex quantum circuit requires calculating approximate circuits of a large unitary matrix. Unfortunately, the calculation of how mathematically approximate a circuit is to its original circuit is computationally infeasible for quantum programs beyond four-five qubits. \sol{} employs circuit partitioning to make the problem tractable: \sol{} applies approximations on small-size partitions (blocks) and then, combines these approximate blocks to produce an approximate full circuit for the original circuit. While this is the only way to produce approximations for large circuits, it poses two primary challenges.

(1) First, when we combine multiple approximate circuit blocks to produce a full approximate circuit, there is no existing knowledge or mechanism to understand how these approximations interact and affect the overall quality of the full approximate circuit. \textit{\sol{} overcomes this hurdle by providing a theoretical proof to guarantee that the quality of the approximation of the full circuit can be bounded by controlling the quality of the approximations of blocks. This allows \sol{} to combine the approximate blocks and guarantee that the overall approximation is of an acceptable quality.}

(2) The second challenge is that the quality of a circuit approximation is expressed as the difference between the unitary matrix of the original circuit and unitary matrix generated after approximation; this measure of approximation does not have strict ordering or direct analytical relationship with the output fidelity. Thus, different ``similar'' approximate circuits can potentially produce different program outputs (and hence, output fidelities). \textit{\sol{} overcomes this challenge by devising a novel apriori selection and combination strategy to indirectly control the overall output fidelity by choosing a set of ``dissimilar'' approximate circuits that have low CNOT gate counts. \sol{}'s method is effective in achieving higher output fidelity by controlling the approximations for individual circuits.}\vspace{4mm}

\noindent\textbf{The main contributions of this work are as following:}

\begin{itemize}

\item \sol{} leverages the approach of circuit synthesis, a technique to generate a mathematically equivalent circuit for a given quantum circuit~\cite{shende2006synthesis,iten2016quantum,martinez2016compiling,iten2019introduction,davis2020towards}, and approximates it to generate low-CNOT-gate-count circuits that deliver a large reduction in noise.

\item \sol{} defines an apriori approximation selection criterion that enables it to select approximations with fewer CNOT gates in a manner that the effect of the approximations does not manifest in the output of the circuit.

\item We provide a theoretical proof to bound the distance of the approximations from the original circuit, which enables \sol{} to partition the circuit for synthesis and scale up. 

\item \sol{}'s evaluation demonstrates a 30-80\% reduction in the CNOT gate count of circuit while ensuring that the output does not deviate from the output of the original circuit on an ideal quantum system. On a noisy system, \sol{} reduces the output error by up 30\% points.

\item Using real materials simulation algorithms like TFIM and Heisenberg, \sol{} demonstrates its ability to track algorithm-specific output even on existing quantum computers due to careful selection of low-error approximations.

\end{itemize}
\begin{figure*}[t]
    \centering
    \includegraphics[scale=0.49]{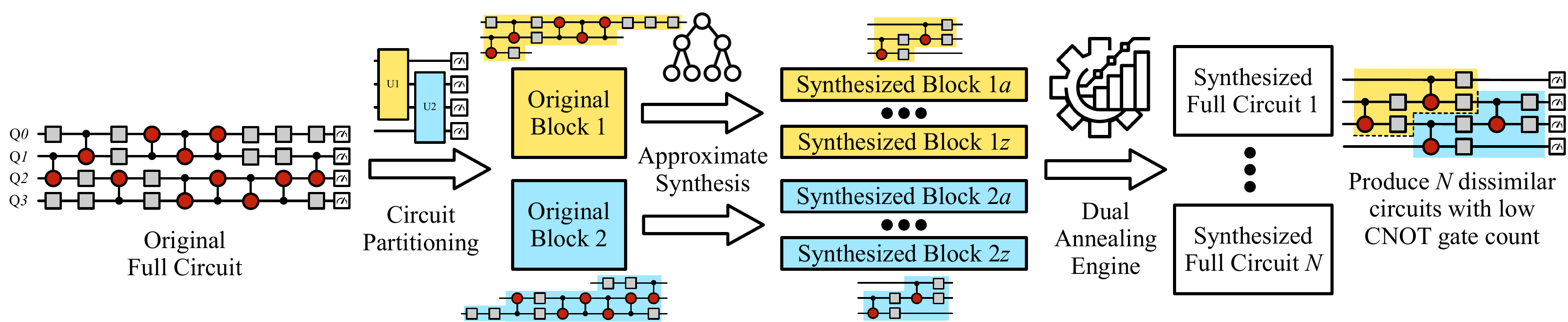}
    \caption{\sol's methodology of partitioning the circuit into smaller blocks, generating multiple approximate solutions for each block, and selecting ``dissimilar'' circuits from the block search space while minimizing the CNOT gate count.}
    \label{fig:end_to_end_cart}
\end{figure*}

\section{\sol{}-Relevant Terms and Definitions}
\label{sec:terms}

Before we present our technique in Sec.~\ref{sec:solution}, below we introduce some terms and definitions that are used in this paper.

A quantum algorithm (or transformation) is a procedure (or computation) that takes a system from an initial state into a final state, as prescribed by its developer. There are multiple formalism spaces in which we can reason about quantum program behavior: 1) process metrics~\cite{kitaev2002classical}, and 2) domain-specific or standard output metrics~\cite{gilchrist2005distance,breuer2009measure,nielsen2010frontmatter}. 

First, at the fundamental level, each program is represented by a unitary matrix $U$. For a $n$-qubit system in initial state $\ket{\psi}$, the effect of the program is computed as $U \ket{\psi}$, where $U$ is a $2^n\times2^n$ unitary matrix. In general, synthesis algorithms use norms to assess the solution quality, and their goal is to minimize $\norm{U - U'}$, where $U$ is the unitary that describes the transformation and $U'$ is the computed solution. This norm is referred to as \textbf{process distance}. Intuitively, process distance metrics compare how alike are two mathematical representations of a computation. While early synthesis algorithms used to employ the diamond norm~\cite{kitaev2002classical}, more recent works use the Hilbert-Schmidt (HS) inner product due to its lower computational overhead~\cite{khatri2019quantum,davis2020towards,kliuchnikov2014asymptotically}.

The Hilbert-Schmidt inner-product is defined as $\Trace(U^{\dag} U')$. This metric ranges from $0$ to $2^n$, and the closer the value is to $0$, the higher the process distance of the two unitary matrices. To make the process distance value range from $0$ to $1$ and to make it so that the closer the value is to $0$, the smaller the distance (``zero'' distance), typically the HS distance is transformed as $\langle U, U' \rangle_{HS} = \sqrt{1 - \frac{\norm{\Trace(U^{\dag}U')}^2}{N^2}}$, where $N=2^n$. We refer to this metric as the \textbf{Hilbert-Schmidt or HS distance} and use it for our approximate synthesis technique.

Second, comparing two quantum programs requires examining and reasoning about their output after the circuit is measured. Given two  probability amplitudes of the output of two quantum algorithms), there are multiple distances used to asses their similarity depending on the use case of the algorithm. For example, for the TFIM and Heisenberg algorithms shown in Fig.~\ref{fig:tfim_heis_mot} the average and staggered magnetization rates are calculated from their output probabilities. However, in general, a common distance measure across all algorithms is required to compare the probability amplitudes of the produced distribution with the expected distribution.

A probability distribution distance metric that is primarily used to measure the \textbf{output distance} is the \textbf{Total Variational Distance} (TVD). For an $n$-qubit circuit, which has $N=2^n$ output states, the TVD can be calculated as $\frac{1}{2}\sum_{k=1}^{k=N}\abs{p(k) - p'(k)}$, where $p(k)$ is the probability of state $k$ with the original circuit and $p'(k)$ is the probability of state $k$ with the synthesized circuit. Another metric is the \textbf{Jensen-Shannon Divergence} (JSD), calculated as $\sqrt{\frac{1}{2}\big{[}D(p\parallel m) + D(p'\parallel m)]}$, where $m$ is the pointwise mean of $p$ and $p'$ and $D$ is the Kullback–Leibler divergence calculated as $\sum_{k}^{} q(k) log(\frac{q(k)}{r(k)})$ for two probability distributions $q$ and $r$. Both metrics have a value between 0 and 1, with 0 being the best (lowest distance). We use these two metrics to evaluate the output distance for all algorithms.

Next, we present \sol{}, an approximate synthesis technique to reduce the CNOT gate count of quantum algorithms.

\section{Design and Implementation of \sol{}}
\label{sec:solution}

In this section, we provide an in-depth description of how \sol{} addresses the problem of reducing the number of CNOT gates in a circuit. We begin by providing an overview of the end-to-end design of \sol{}.

\subsection{Overview of \sol{} Design}

As shown in Fig.~\ref{fig:end_to_end_cart}, \sol{} first partitions the large circuit into multiple smaller circuits called ``blocks'' to ensure a scalable solution. Next, it generates approximate circuit solutions with low CNOT gate counts for individual blocks using approximate synthesis. Approximate synthesis is a procedure whereby an approximate circuit can be produced by reducing the process distance between the unitary matrices of the approximate and original circuits. Once the synthesized blocks are generated for each block, they are combined to form the approximate full circuit, which now has a lower CNOT gate count than the original circuit.

However, because the circuit is approximate, the output of the circuit can be different from the original circuit. \textit{To overcome this issue, \sol{}'s makes use of its key insight: multiple different approximate full circuits can be generated because the block instances can be combined in different ways.} \sol{} uses a dual annealing engine to search the block instance search space in a manner that it selects multiple low-CNOT-gate-count full circuit approximations that are mathematically ``dissimilar'' to each other. This ensures that when the outputs of the approximations are averaged, it produces the same output as the original full circuit, but by using circuits that have fewer CNOT gates. We also provide a theoretical proof that bounds the full circuit's approximation distance, without the need to calculate it directly as it is computationally infeasible, to avoid coarse approximations. 

Next, we describe each of these steps in detail, starting with a description of circuit synthesis.

\subsection{Overview and Challenges of Circuit Synthesis}

Quantum circuit synthesis is the technique used to find a circuit for a quantum algorithm that is mathematically close (referred to in this paper as ``exact'' synthesis) to the original algorithm circuit. Recall that an $n$-qubit quantum algorithm can be represented as an $N\times N$ unitary matrix, where $N = 2^n$. This unitary matrix can be calculated by taking a product of all the operations run by a quantum algorithm. For example, if an algorithm runs $K$ operations one after another, and the $k^{th}$ operation is represented using the unitary $U_k$, the unitary of the entire algorithm can be calculated as $U = U_K U_{K-1}\dots U_2 U_1$.

\begin{figure}[t]
    \centering
    \includegraphics[scale=0.75]{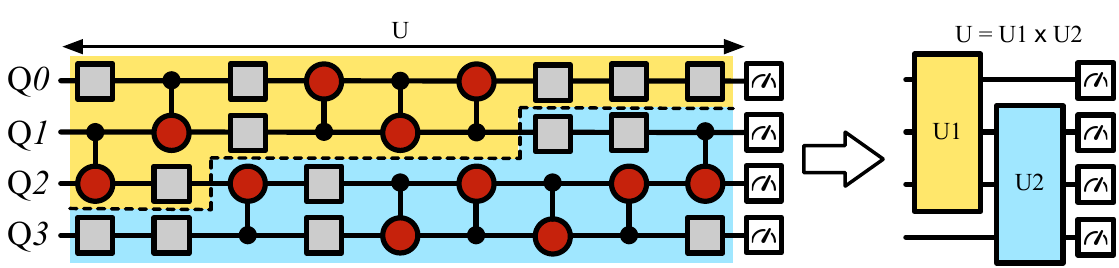}
    \caption{Example of partitioning a circuit (represented using unitaries $U_1$ and $U_2$). Operations are applied in order from left to right. Isolated squares represent one-qubit rotation gates, circles connected to another qubit represent CNOT gates, and the squares at the end represent measurement gates.}
    \label{fig:part_cart}
\end{figure}

Synthesis is the process of finding a circuit with unitary $U'$ (with fewer CNOT gates than the original) for $U$, such that the process distance between the two is minimized. It is non-trivial to construct an optimal $U'$ (in terms of CNOT gates) in an analytical or rule-based manner for large circuits~\cite{tucci2005introduction,iten2019introduction}. For this reason, previous works have proposed numerical optimization solutions~\cite{davis2020towards,younis2021qfast,smith2021leap}. These works attempt to construct the synthesized circuit one layer at a time (a layer typically consists of a combination of one-qubit rotation gates and two-qubit CNOT gates). Once a layer is embedded onto the synthesized circuit, numerical optimization methods are used to optimize the rotation angles such that the process distance between the unitary of the synthesized circuit ($U'$) and the original circuit ($U$) is minimized. If the distance is within a certain acceptable threshold, the solution is accepted. If an acceptable solution is not found, another layer of gates is added to the synthesized circuit and again the numerical optimization is performed. Note that every time a layer is added, the optimizer has increased degrees of freedom due to the more rotation angles, getting the synthesized circuit's unitary closer to the original circuit's unitary. The Hilbert-Schmidt process distance is widely used to calculate whether the process distance between $U'$ and $U$ is within the acceptable threshold of $\epsilon$~\cite{khatri2019quantum,davis2020towards,kliuchnikov2014asymptotically}: $\sqrt{1 - \frac{\norm{\Trace{}(U^\dagger U')}^2}{N^2}} < \epsilon$. 

\textbf{However, the synthesis technique is not scalable with increase in circuit size as the unitary scales exponentially with the number of qubits.} This makes the calculation of the process distance during each iteration of numerical optimization exponentially slower with increase in circuit size. For example, a 20-qubit circuit requires $2^{20\times 2}$ inner products to calculate the process distance each time. In fact, it is difficult to calculate the unitary of the entire circuit in the first place because it requires multiplying large quantum operation unitaries. Thus, it is infeasible to perform synthesis on a full circuit unitary for large circuits. The circuit can be divided into manageable chunks before synthesis can be performed. This is the first step of the \sol{} procedure.

\subsection{STEP 1: Partitioning Large Circuits}

A solution to the problem of scaling the circuit synthesis approach is to partition the circuit into blocks of smaller sizes~\cite{bqskit}. Fig.~\ref{fig:part_cart} shows how this can be achieved for an example circuit of four qubits. Assuming that we can computationally synthesize circuits that are up to three qubits in size, if we partition the four-qubit circuit into two blocks of three-qubits, we can synthesize the two blocks separately to generate circuits equivalent to their corresponding unitaries: $U_1$ and $U_2$. The blocks are formed such that there are no connections in terms of two-qubit CNOT gates between the two blocks. Otherwise, they may be entangled and cannot be synthesized separately.

In terms of our example circuit, the unitary $U_1$ is synthesized such that $1 - \frac{\norm{\Trace{}(U^\dagger_1 U'_1)}}{N_1} < \epsilon_{1}$ and the matrix $U_2$ is synthesized such that $1 - \frac{\norm{\Trace{}(U^\dagger_2 U'_2)}}{N_2} < \epsilon_{2}$. When the synthesized circuits are obtained for $U_1$ and $U_2$, they are put together to form the full synthesized circuit. Note that this approach is scalable because we limit the block size to what is computationally possible. Synthesizing in this manner does not require generating the target unitary for the full circuit, nor does it require calculating the process distance for the entire circuit's unitary. However, this approach also has several challenges, which we discuss in the following section.

\subsection{Challenges with Exact Synthesis of Blocks}

\begin{figure}[t]
    \centering
    \includegraphics[scale=0.56]{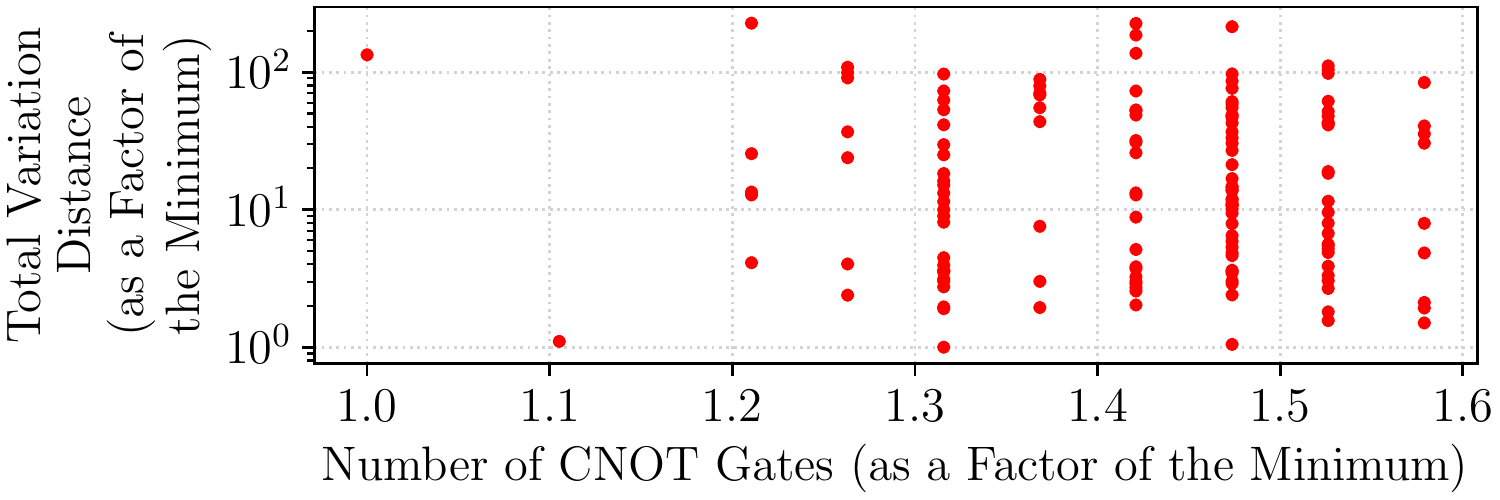}
    \caption{Relationship between the number of CNOT gates and the TVD for several exactly synthesized solutions of a four-qubit Variational Quantum Eigensolver (VQE) circuit.}
    \label{fig:rmd_vs_tvd}
\end{figure}

An exact solution that minimizes the process distance might not necessarily minimize: 1) the CNOT gate count, and 2) the output distance: how far the output of the synthesized circuit is from the output of the original circuit. As for 1), exact synthesis solutions provide a small reduction in the number of CNOT gates due to the strict process distance threshold requirements (more layers have to be added to the circuit during synthesis to allow for more degrees of freedom). 

With regard to 2), the output distance cannot be used as a metric of mathematical exactness during synthesis because the state of a quantum system can only be represented using its unitary. Moreover, when the circuit is partitioned, each block cannot be optimized using an output distance metric because individual blocks have no output of their own. Only the full circuit has an output and it can only be measured after execution. Thus, while the process distance is required for synthesis, the output distance comes into play when the output of the algorithm needs to be interpreted. 

However, circuits with similar process distances can have different output distances and CNOT gate counts. Fig.~\ref{fig:rmd_vs_tvd} demonstrates the relationship between the number of CNOT gates and the output distance (TVD) for several exactly synthesized solutions of a four-qubit Variational Quantum Eigensolver (VQE) circuit. All solutions have a similar process distance of less than $10^{-5}$ (exact solution threshold) and yet have TVDs in a large range. The solution with the minimum number of CNOT gates has one of the highest TVDs, while the solution with only 10\% more CNOT gates than the minimum (1.1 factor) has a lower TVD. This small-scale example shows that it is not always advisable to select the exact solution with the fewest number of CNOT gates. Thus, \sol{} breaks away from the notion of having an exact synthesized solution by designing an approximate synthesis approach.




\subsection{STEP 2: Generate Approximate Circuits for Blocks}

\begin{figure}[t]
    \centering
    \includegraphics[scale=0.58]{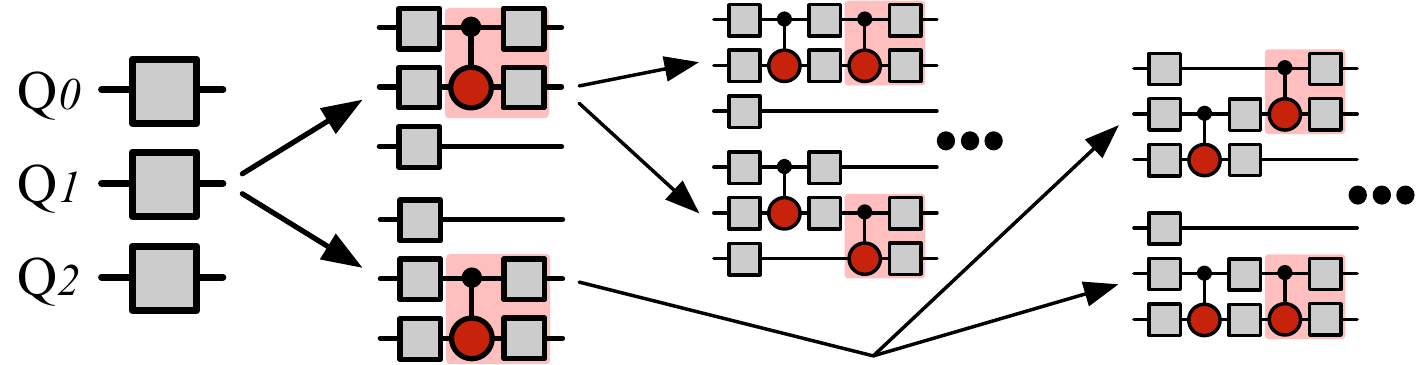}
    \caption{Leap compiler builds the circuit layer-by-layer.}
    \label{fig:leap_compiler}
\end{figure}

\sol{} employs an approximate synthesis procedure that generates multiple low-CNOT-gate-count circuit approximations in a manner that minimizes the output distance between the synthesized circuits and the original circuit of a block.

The key to achieving this reduction is realizing that different types of circuits can be synthesized such that they have process distance below a certain threshold. Recall that synthesis is performed using numerical optimization building layer-by-layer. The qubits that these layers are placed on and the rotation angles that are assigned to them during the process distance minimization procedure affects the final synthesized circuit that is produced. Multiple different approximate circuits can be produced by varying these factors.

To achieve this, \sol{} modifies the Leap compiler~\cite{smith2021leap} to return the best $M$ circuits with the lowest process distance at each layer of the compiler tree. The compiler constructs a circuit tree one layer at a time as shown in Fig.~\ref{fig:leap_compiler}. Each layer consists of a CNOT gate between two qubits followed by two rotation gates on both the qubits. It attempts this layer on all allowed two-qubit combinations and optimizes the rotation angles to minimize the unitary process distance. As more layers are added, the tree expands due to the increased number of layer permutations. Every few layers, it picks the branch with the least process distance and starts reconstructing the tree from there to reduce the number of optimization evaluations. \sol{} uses the compiler to generate multiple approximations at each layer of the tree (a tree layer roughly corresponds to one CNOT gate). This enables the generation of approximate circuits of different qualities in terms of process distances for each block at different CNOT gate counts. All approximate solutions are generated until the tree exceeds the CNOT gate count of the original circuit. As more layers are constructed and the tree becomes deeper, the process distance decreases. However, because we would like to have multiple solutions for a block, including the ones with higher CNOT gate counts than the minimum CNOT gate count solution, we collect all solutions of different CNOT gate counts.


\subsection{STEP 3: Putting Together Approximate Blocks}

\begin{figure}[t]
    \centering
    \includegraphics[scale=0.32]{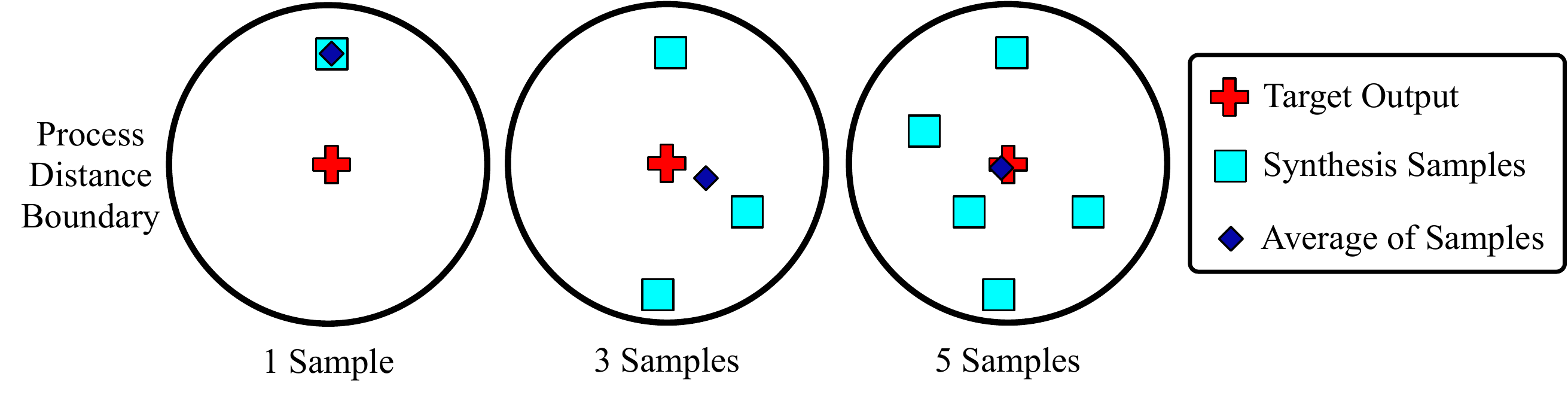}
    \caption{Visual representation of how averaging over multiple approximations in the can produce a similar output as the original circuit (with the average having a low output distance).}
    \label{fig:out_dist_cart}
\end{figure}

How can these low-CNOT-gate-count approximations be used to reduce the output distance? Fig.~\ref{fig:out_dist_cart} demonstrates a visual example of how selecting multiple synthesized circuits can help ensure that the output distance is reduced. The figure is read from left to right, with the first circle showing when one approximate circuit sample is used and the last circle showing when six samples are used. The red cross shows the output of the original circuit in Hilbert space, the boundary circle around it demarcates the boundary of the process distance threshold used for approximate synthesis (this threshold is larger than used with exact synthesis so as to generate low-CNOT-gate-count solutions). The blue squares show the output samples of the approximate synthesized circuit and the dark blue diamond shows the average of the samples. We make two points:

1) If only the circuit with the fewest number of CNOT gates is selected, it can result in a high output distance as shown in the first circle. While the circuit with the lowest CNOT gate count might also coincidentally have a low output distance for some algorithms, this cannot be established analytically as it assumes knowing the ground truth output. \textit{Instead, averaging over $M$ low-CNOT-gate-count circuits can help us regulate and control the output distance with more robustness than simply choosing one circuit.} The trade-off here is between the CNOT gate count and the output distance. If multiple circuits are not selected, we do not have a way to control and reduce the output error. On the other hand, if the CNOT gate count is high, it defeats the purpose of synthesis.

2) It is also not sound to just have many random approximations. If the approximations are mathematically similar (e.g., if the six samples in the third circle were in the same region of the circle), their output cannot average out to reduce the output distance. Thus, \sol{} must ensure that the approximations are ``dissimilar,'' while also having a low CNOT count.

\begin{algorithm}[t]
\caption{Dual annealing engine's objective function.}
\label{alg:gen_app}
\begin{algorithmic}[1]
\State $O \gets$ Original circuit
\State $A \gets$ Approximate circuit to score
\State $S \gets$ Already selected approximations
\State $c_{norm} \gets$ (CNOT count of $A$) $/$ (CNOT count of $O$)
\State $\epsilon \gets$ Process distance threshold
\If{$\langle A, O \rangle_{HS} > \epsilon$} \Comment{If the threshold is breached}
    \State \Return 1.0
\ElsIf{$S == \{\varnothing\}$} \Comment{If this is the first sample}
    \State \Return $c_{norm}$
\Else
    \State $m = 0$ \Comment{Fraction of similar samples}
    \For{$s \in S$}
        \State $m = m + \langle A, s\rangle_{HS} \ \le \ \max\{\langle A, O\rangle_{HS}, \langle O, s\rangle_{HS}\}$
    \EndFor
    \State $m = m / \text{size}(S)$
    \State \Return $\frac{1}{2} \times m + \frac{1}{2} \times c_{norm}$
\EndIf
\end{algorithmic}
\end{algorithm}

\sol{} achieves this balance by using a dual-annealing-based minimization algorithm~\cite{sahin1998dual} shown in Algorithm~\ref{alg:gen_app} to minimize an objective function that places equal weight on CNOT-gate count and the dissimilarity of the approximations: $\min f = \frac{1}{2}\times\text{CNOT Count} + \frac{1}{2}\times\text{Approx. Dissimilarity}$

The CNOT count in this objective function is simply the normalized CNOT gate count of the approximation compared to the original circuit. The approximation dissimilarity is calculated as the fraction of already selected circuit samples with similarity to the new sample. Consider two approximate circuit samples, $S_1$ and $S_2$. If the process distance between the two samples ($\langle S_1, S_2\rangle_{HS}$) is less than the maximum of their process distances to the original circuit ($\max\{\langle S_1, O\rangle_{HS}, \langle O, S_2\rangle_{HS}\}$), then the two samples are considered similar: $\langle S_1, S_2\rangle_{HS} \ \le \ \max\{\langle S_1, O\rangle_{HS}, \langle O, S_2\rangle_{HS}\}$. Intuitively, in the visualization shown in Fig.~\ref{fig:out_dist_cart}, this means that both samples are in the same region of the circle. If the process distance between the two is greater than the maximum of their process distances to the original circuit, then the two circuits are on the opposite side of the circuit; thus, their output can be averaged out. The first sample has an approximation dissimilarity of zero (since there are no already selected circuits), and the objective function will select the approximate circuit with the lowest CNOT-gate count. As more samples are selected, the approximation dissimilarity gains more significance as it becomes increasingly difficult to find dissimilar approximations. However, the equal weight on CNOT count ensures that approximations that have too many CNOTs are not selected.

While this objective function works well for non-partitioned circuits, it requires a tweak to accommodate large partitioned circuits. Calculating if two full circuit approximations (constructed by putting together block approximations) are similar should not require the computationally infeasible use of the full circuit unitaries. Instead, \sol{} uses the metric ``fraction of all circuit blocks that are similar'' in the objective function as it is a scalable alternative that works well in practice. As an instance, for two approximations of a full circuit consisting of ten blocks, if three of the blocks are mathematically similar, the two approximations receive a similarity score of $0.3$. Next, we discuss a major challenge when approximating large partitioned circuits.

\subsection{A Challenge of Approximating Full Circuits}

While for individual blocks it can be ensured that the approximations are not too coarse by eliminating ones with a high output distance (Lines 6-7 in Algorithm~\ref{alg:gen_app}), a major challenge of partitioned synthesis is to ensure that when the approximate blocks are put back together, the process distance of the full circuit approximation is not violated. Blocks cannot be combined without the knowledge of how their process distances accumulate. For example, the process distance may compound multiplicatively when the blocks are combined to form the full circuit. Without having a theoretical bound on the process distance of the full circuit, the process distance thresholds of its blocks may have to be kept unnecessarily small to be on the safe side. This means that the synthesized blocks likely end up being longer than they need to be as more layers are typically required during the numerical optimization process if the distance threshold is very small. Overcoming this problem can help us synthesize shorter circuits with fewer CNOT operations. To this end, next we provide a theoretical proof to bound the full circuit process distance based on the process distances of its blocks without the need to directly calculate the process distance of the full circuit. This will help us ensure that the full circuit approximations that are too coarse can be eliminated during the dual annealing minimization procedure.

\subsection{Theoretical Upper Bound on Process Distance}

We prove the theoretical upper bound for a circuit partitioned into two blocks without any loss of generality (e.g., the one shown in Fig.~\ref{fig:part_cart}), and it can then be extended to a circuit partitioned into $K$ blocks.

We want the process distance of the unitary $U$ of the full circuit to be bounded without performing synthesis on the full circuit due to the lack of scalability: $\sqrt{1 - \frac{\norm{\Trace{}(U^\dagger U')}^2}{N^2}} < \epsilon$. $U$ is an $N \times N$ matrix, where $N$ is $2^n$, where $n$ is the number of qubits in the full circuit. We choose the $\sqrt{1 - \frac{\norm{\Trace{}(U^\dagger U')}^2}{N^2}}$ metric for process distance as it is a reasonable metric to measure unitary equivalency, it is computationally efficient, and it gives us the ability to prove a bound on it without calculating it directly. The $\epsilon$ bound needs to be derived based on the process distances of its circuit blocks. Recall that the partitioned blocks are small enough to be efficiently synthesizable and therefore, have known process distances. The example circuit has two blocks and these blocks have the below two process distance bounds by construction (synthesis is performed in a manner that ensures that these bounds are met).
    \begin{gather+}[0.9]
    \sqrt{1 - \frac{\norm{\Trace{}(U^\dagger_1 U'_1)}^2}{N_1^2}} \le \epsilon_{1} \ \ \ , \ \ \ 
    \sqrt{1 - \frac{\norm{\Trace{}(U^\dagger_2 U'_2)}^2}{N_2^2}} \le \epsilon_{2}
    \label{eq:thames}
    \end{gather+}

Here, $N_1$ and $N_2$ are the dimensions of $U_1$ and $U_2$, respectively. For the example circuit, we have $U = (I \otimes U_2) (U_1 \otimes I) = U_{I2} U_{1I}$. The notation $U_{1I}$ refers to the unitary $(U_1 \otimes I)$, representing the Kronecker product of the $U_1$ operation with the identity operation on the remaining qubits (no operation can be represented as the identity operation). Similarly, $U_{I2}$ refers to the unitary $(I \otimes U_2)$. The post-synthesis approximations of these unitaries can be represented as $U' = (I \otimes U'_2) (U'_1 \otimes I) = U'_{I2} U'_{1I}$. Also, $N = N_I \times N_1 = N_2 \times N_I$.

As a first step, we prove the process distance bound when a unitary representing a partitioned block is extended to the remaining qubits in the full circuit, i.e., we determine the process distance of $U_1 \otimes I$ matrix given the process distance of the $U_1$ matrix. We begin by rearranging the terms in Eq.~\ref{eq:thames} to isolate for $\norm{\Trace{}(U^\dagger_1 U'_1)}$, as is shown below.

    \begin{gather+}[0.9]
        \sqrt{1 - \frac{\norm{\Trace{}(U^\dagger_1 U'_1)}^2}{N_1^2}} \le \epsilon_1 \Rightarrow \frac{\norm{\Trace{}(U^\dagger_1 U'_1)}^2}{N_1^2} \ge 1 - \epsilon^2_1 \\
        \Rightarrow \norm{\Trace{}(U^\dagger_1 U'_1)}^2 \ge N_1^2(1 - \epsilon^2_1) \Rightarrow \norm{\Trace{}(U^\dagger_1 U'_1)} \ge N_1\sqrt{1 - \epsilon^2_1}
   \label{eq:nile}
   \end{gather+}

Next, we show how $\Trace{}(U^\dagger_{1I}U'_{1I})$ is related to $\Trace{}(U^\dagger_1U'_1)$:
    \begin{gather+}[0.9]
        \Trace{}(U^\dagger_{1I}U'_{1I}) = \Trace{}[(U_1 \otimes I)^\dagger(U'_1 \otimes I)] = \Trace{}[(U^\dagger_1 \otimes I^\dagger)(U'_1 \otimes I)] \\
         = \Trace{}[(U^\dagger_1 \otimes I)(U'_1 \otimes I)] = \Trace{}(U^\dagger_1U'_1 \otimes II) = \Trace{}(U^\dagger_1U'_1 \otimes I) \\
         = \Trace{}(U^\dagger_1U'_1)\text{tr(I)} = \Trace{}(U^\dagger_1U'_1)N_I
    \label{eq:amazon}
    \end{gather+}
Substituting Eq.~\ref{eq:nile} into Eq.~\ref{eq:amazon}, we get that $\norm{\Trace{}(U^\dagger_{1I} U'_{1I})} \ge N\sqrt{1 - \epsilon^2_1}$, as shown below.
    \begin{gather+}[0.9]
        \norm{\Trace{}(U^\dagger_{1I} U'_{1I})} = \norm{\Trace{}((U^\dagger_1U'_1)N_I} = \norm{\Trace{}((U^\dagger_1U'_1)}N_I \\
        \ge N_1\sqrt{1 - \epsilon^2_1}N_I = N_1 N_I\sqrt{1 - \epsilon^2_1} = N\sqrt{1 - \epsilon^2_1}
    \label{eq:neponset}
    \end{gather+}
If we rearrange the terms in Eq.~\ref{eq:neponset} (similar to the rearranging in Eq.~\ref{eq:nile}, but in reverse order), we get $\sqrt{1 - \frac{\norm{\Trace{}(U^\dagger_{1I} U'_{1I})}^2}{N^2}} \le \epsilon_1$. Thus, the process distance of a block unitary that does not span the full size (number of qubits) of a circuit remains bounded by the same threshold when the unitary is extended to the size of the circuit. Therefore, using a similar procedure, we can also show that for the second block, $U_2$, $\sqrt{1 - \frac{\norm{\Trace{}(U^\dagger_{I2} U'_{I2})}^2}{N^2}} \le \epsilon_2$.


We now have the tools to bound the process distance of the full circuit. Recall that the process distance for the full circuit is $\sqrt{1 - \frac{\norm{\Trace{}(U^\dagger U')}^2}{N^2}}$. Substituting $U = U_{I2} U_{1I}$, we obtain the following:
    \begin{gather+}[0.9]
    \sqrt{1 - \frac{\norm{\Trace{}(U^\dagger U')}^2}{N^2}} = \sqrt{1 - \frac{\norm{\Trace{}[(U_{I2} U_{1I})^\dagger(U'_{I2} U'_{1I})]}^2}{N^2}} \\
    = \sqrt{1 - \frac{\norm{\Trace{}(U^\dagger_{1I} U^\dagger_{I2} U'_{I2} U'_{1I})}^2}{N^2}} 
    = \sqrt{1 - \frac{\norm{\Trace{}[(U'_{1I} U^\dagger_{1I})(U^\dagger_{I2} U'_{I2})]}^2}{N^2}}
    \label{eq:hudson}
    \end{gather+}
Using the inequality proven by Wang and Zhang~\cite{wang1994trace} and using the above derived bounds for process distances of $U_{1I}$ and $U_{I2}$, the following relationship is obtained:
    \begin{gather+}[0.9]
    \sqrt{1 - \frac{\norm{\Trace{}[(U'_{1I} U^\dagger_{1I})(U^\dagger_{I2} U'_{I2})]}^2}{N^2}} \\
    \le \sqrt{1 - \frac{\norm{\Trace{}(U'_{1I} U^\dagger_{1I})}^2}{N^2}} + \sqrt{1 - \frac{\norm{\Trace{}(U^\dagger_{2I} U'_{2I})}^2}{N^2}} \\
    \le \sqrt{1 - \frac{\norm{\Trace{}(U^\dagger_{1I} U'_{1I})}^2}{N^2}} + \sqrt{1 - \frac{\norm{\Trace{}(U^\dagger_{2I} U'_{2I})}^2}{N^2}} \\
    \le \epsilon_{1} + \epsilon_2
    \label{eq:danube}
    \end{gather+}
Combining Eq.~\ref{eq:hudson} and Eq.~\ref{eq:danube}, we get $\sqrt{1 - \frac{\norm{\Trace{}(U^\dagger U')}^2}{N^2}} \le \epsilon_1 + \epsilon_2$. This proof can be extended to circuits partitioned into $K$ blocks by providing the proof for two blocks at a time, combining the two into a single unitary and providing the proof again for the two unitaries (combined unitary and the third unitary), and so on and so forth, iteratively. \textit{Therefore, for a circuit partitioned into $K$ blocks, we have that $\sqrt{1 - \frac{\norm{\Trace{}(U^\dagger U')}^2}{N^2}} \le \displaystyle\sum_{k=1}^{k=K}\epsilon_k$. The process distance of the full circuit is theoretically upper bounded by the sum of the process distances of all of its partitioned blocks.}

\begin{figure}[t]
    \centering
    \includegraphics[scale=0.48]{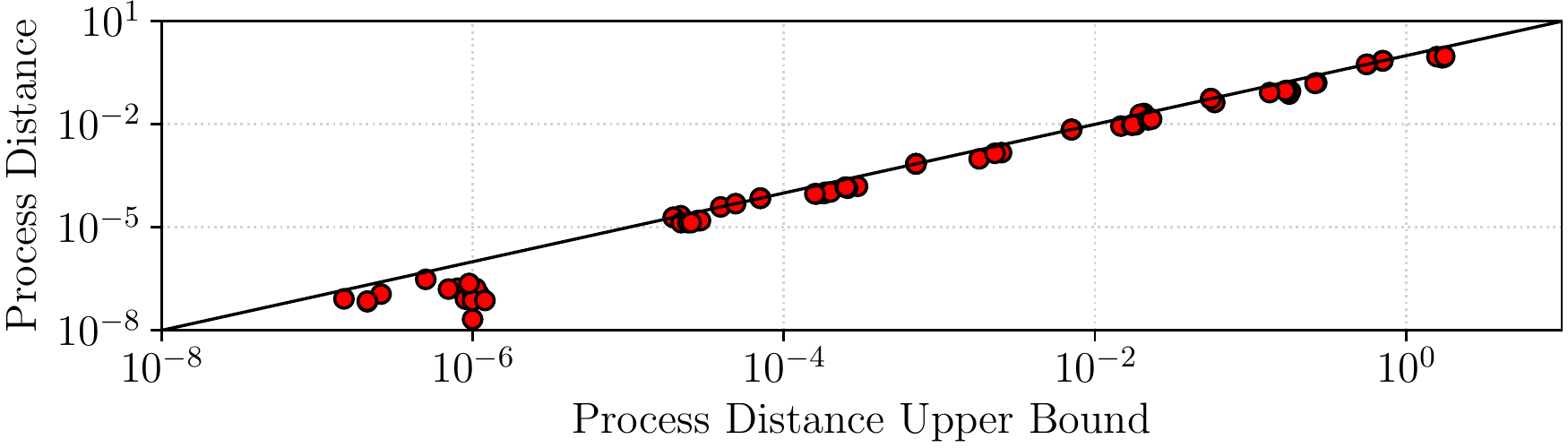}
    \caption{The theoretical process distance upper bound successfully bounds the process distance in practice.}
    \label{fig:hs_up}
\end{figure}

While the relationship between the derived process distance upper bound and the actual process distance cannot be directly/theoretically proven as it algorithm specific and varies depending on how the algorithm circuit is constructed and partitioned, we demonstrate this relationship using real algorithm examples. Fig.~\ref{fig:hs_up} shows the relationship between the process distance upper bound and the actual process distance for different algorithms. The results indicate that the derived upper bound is respected across all samples and a relatively tight bound is obtained for different algorithms and process distance values. Therefore, the upper bound enables us to confidently bound the process distance of the full circuit (without the need to calculate it directly) simply by ensuring that its partitioned blocks are combined by the dual annealing engine in a manner that the sum of their process distances is within an acceptable threshold.



\textbf{\newline Putting it togerher.} {In summary, \sol{} enables reduction in the CNOT gate count by partitioning the circuit into smaller blocks and generating multiple approximate circuits for the blocks. It then puts back together the block approximations in a manner that reduces the CNOT gate count as well as generates dissimilar approximations to reduce the output distance using its dual annealing engine. It is aided by the theoretical upper bound to eliminate coarse approximations in a scalable manner}. In the next section, we evaluate the effectiveness of \sol{} for different algorithms after providing details about the experimental setup and methodology.


\section{Evaluation}
\label{sec:evaluation}

\subsection{Experimental Setup and Methodology}

\noindent\textbf{Comparative Techniques.} We compare against the original circuit as a baseline circuit (referred to as the \textbf{Baseline}) in terms of the CNOT gate count, ground truth output, as well as the error observed in a noisy environment. We also compare to the circuit generated when all the Qiskit compiler optimizations are applied to this Baseline circuit. These compiler optimizations are simply referred to as \textbf{Qiskit}. When these compiler passes are applied to the approximate circuits produced by \sol{}, the results generated by the produced circuits are referred to as \textbf{\sol{} + Qiskit}.

\noindent\textbf{\newline Experimental Setup.} The partitioning, approximate synthesis, and dual annealing components of \sol{} are run on our local compute cluster consisting of 2.4 GHz Intel E5-2680 v4 CPUs. A maximum block size of four qubits is used to partition the circuits using the scan partitioner available as part of the open-source BQSKit package~\cite{bqskit}. It forms partitions by traversing the circuit left to right and creating four-qubit blocks. Note that some blocks may be of a smaller size if need be. A maximum block size of four qubits is used as it synthesizes efficiently and yields good results. When an algorithm has multiple blocks, the approximate synthesis step is run on different blocks in parallel on up to ten compute nodes. Leap compiler~\cite{smith2021leap}, which is also a part of the BQSKit package, is modified and used for synthesis as described in Sec.~\ref{sec:solution}. The Python-based SciPy package is used to run the dual annealing engine~\cite{jones2016scipy}. The process distance threshold to eliminate coarse approximations in the annealing engine is set proportional to the number of blocks in the circuit and up to 16 approximations are generated for each algorithm. A balanced weight of 0.5 is used for CNOT gate count and approximation dissimilarity in the objective function that the dual annealing engine minimizes. The ground truth results are obtained by running the Baseline circuit in an ideal quantum simulation environment using the Qiskit unitary simulator~\cite{mckay2018qiskit}, which is part of the Aer package. We perform noisy simulations on circuits up to 16 qubits (it was not possible to run noisy simulations on larger circuits) using the IBMQ QASM simulator available via the IBM quantum experience cloud. A Pauli noise model is used for all the qubits with noise levels of 1\%, 0.5\%, and 0.1\% to simulate how \sol{} will perform for future NISQ computers as the noise level decreases. We run circuits up to five qubits on the IBMQ Manila quantum computer in order to demonstrate how the technique performs on existing quantum computers. We use 8192 experimental trials (maximum allowed) per experiment. Each circuit run on a quantum computer takes 10-12 seconds for all trials.

\begin{table}
    \centering
    \caption{Algorithms and benchmarks used to evaluate \sol{}.}
    \scalebox{0.9}{
    \begin{tabular}{|p{1.5cm}|p{6.5cm}|}
    \hline
    \textbf{Algorithm} & \textbf{Description} \\
    \hline
    Adder & Quantum adder circuit~\cite{cuccaro2004new} \\
    \hline
    Heisenberg & Time-independent Heisenberg Hamiltonian~\cite{bassman2021arqtic}  \\
    \hline
    HLF & Hidden linear function~\cite{bravyi2018quantum} \\
    \hline
    QFT & Quantum Fourier transform~\cite{namias1980fractional} \\
    \hline
    QAOA & Quantum alternating operator ansatz~\cite{farhi2016quantum} \\
    \hline
    Multiplier & Quantum multiplier circuit~\cite{hancock2019cirq} \\
    \hline
    TFIM & Transverse field ising model~\cite{bassman2021arqtic}  \\
    \hline
    VQE & Variational quantum eigensolver~\cite{mcclean2016theory} \\
    \hline
    XY & XY quantum Heisenberg model~\cite{bassman2021arqtic} \\
    \hline
    \end{tabular}}
    \label{tab:algorithms}
\end{table}

\noindent\textbf{\newline Algorithms and Benchmarks.} We use the algorithms and benchmarks listed in Table~\ref{tab:algorithms} to evaluate \sol{}. Adder and Multiplier are standard quantum arithmetic circuits, while QAOA and VQE are quantum variational algorithms. Heisenberg, TFIM, and XY are time-evolving Hamiltonian algorithms for material simulations. The Heisenberg model has non-zero strength for the coupling interaction between nearest neighbor spins for all three axes ($x, y, z$), TFIM does for $z$, and XY does for $x$ and $y$. We evaluate circuits of size 4-32 qubits.

\noindent\textbf{\newline Evaluation Metrics.} In terms of output distance, we use the Total Variation Distance (TVD) and the Jensen-Shannon Divergence (JSD) as defined in Sec.~\ref{sec:terms}. These are general metrics that are typically used to define the output distance across all algorithms. To calculate the output distance from the Baseline for \sol{}, the output probability distributions of all of its approximate circuits are averaged to generate one probability distribution. In addition to these metrics, we also study algorithm-specific output distances as necessary. For example, for TFIM and Heisenberg algorithms, we study the differences in the magnetization at different time steps.

\subsection{Results and Analysis}

\begin{figure}[t]
    \centering
    \includegraphics[scale=0.48]{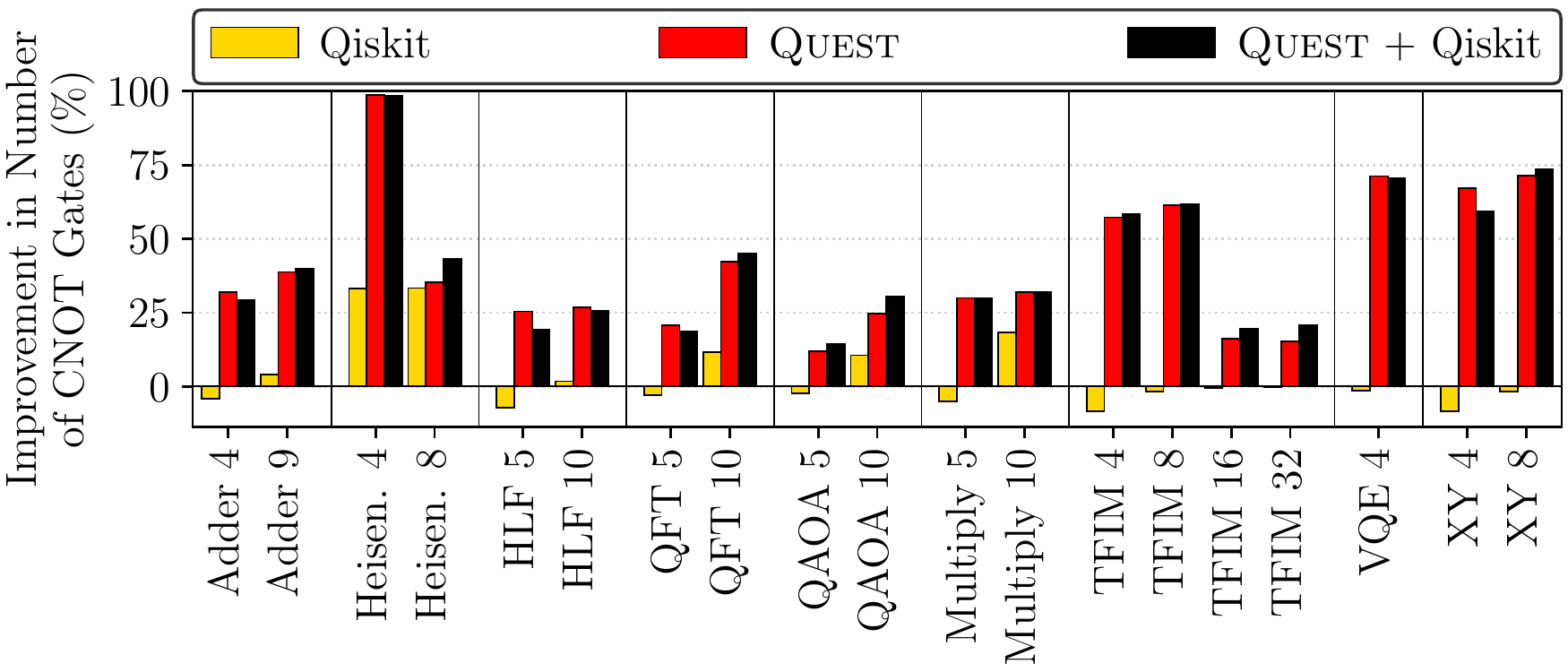}
    \caption{\sol{} reduces the CNOT gate count across all algorithms compared to the Baseline circuits as well as the Qiskit optimizations applied on the Baseline circuits. The number next to the algorithm name indicates the number of qubits.}
    \label{fig:num_cnots}
  \end{figure}

\noindent\textit{\sol{} considerably reduces the CNOT gate count over the Baseline circuit compared to the Qiskit compiler optimizations.} Fig.~\ref{fig:num_cnots} shows the percent improvement, i.e., reduction, in CNOT gate count over the Baseline circuit with Qiskit, \sol{} and \sol{} + Qiskit for different algorithms. \sol{} delivers a reduction in CNOT gate count of 30-80\% for most algorithms, even greater than 80\% for some algorithms such as Heisenberg, which has many CNOT gates. In comparison, the Qiskit compiler optimizations can prove to be less effective depending on the algorithm. For the Heisenberg circuit, Qiskit gives over a 30\% reduction in the CNOT gate count. But for most other circuits the reduction is negligible, even resulting in a slight increase in the CNOT gate count for some algorithms such as HLF and Multiply. In contrast, \sol{} always performs better than Qiskit and never performs worse than the Baseline.

When the Qiskit compiler optimizations are added on top of the approximate circuits produced by \sol{}, there can be a slight improvement or degradation in CNOT count depending on the algorithm. For example, \sol{} + Qiskit performs better than \sol{} for the eight-qubit Heisenberg circuit, but it performs worse for the four-qubit XY circuit. Nonetheless, for most algorithms we observe that it does not diminish the gains of \sol{} and so we use the \sol{} + Qiskit configuration for evaluation results going forward.

\begin{figure}[t]
    \centering
    \subfloat[][Total Variation Distance]{\includegraphics[scale=0.48]{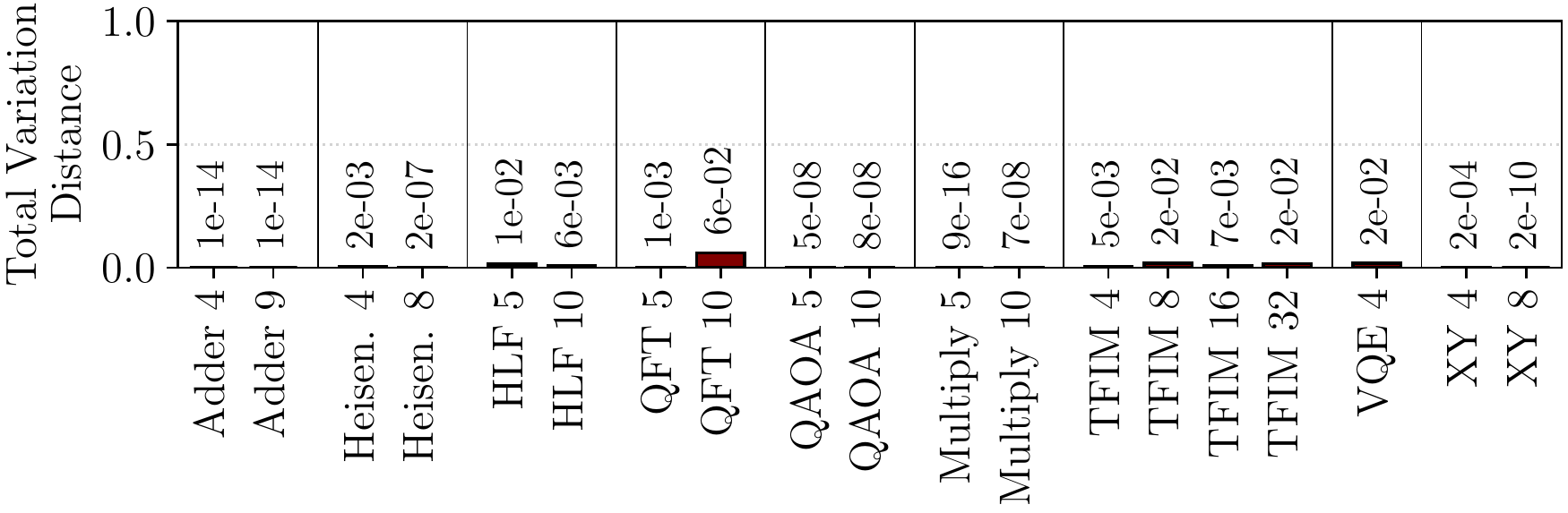}} \vspace{0mm}
    \subfloat[][Jensen-Shannon Divergence]{\includegraphics[scale=0.48]{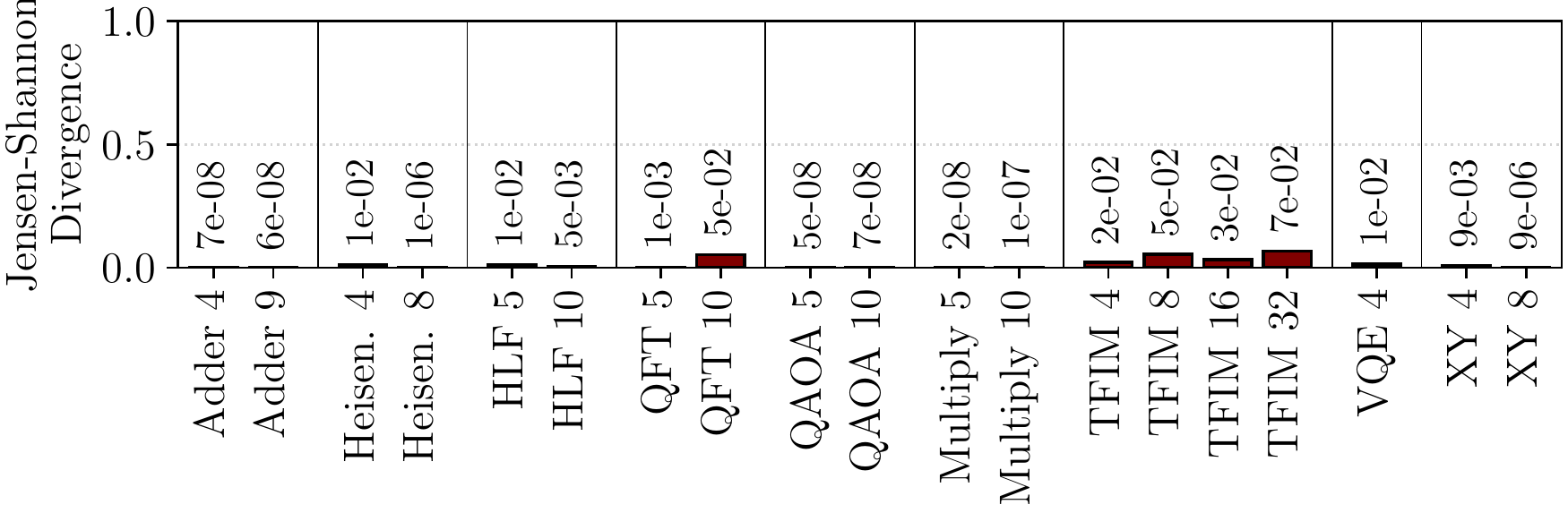}}
    \caption{\sol{} ensures low output distance from the ideal output while delivering a reduction in CNOT gate count.}
    \label{fig:ideal_noise}
  \end{figure}

\noindent\textit{\newline \sol{}'s approximations result in a low output distance even in the ideal quantum computing scenario.} We now evaluate if the approximate circuits produced by \sol{} are resulting in the correct output (close to ground truth results) even in the absence of noise. This can help us understand if the approximate circuits are closely emulating the expected output of the Baseline circuit. Fig.~\ref{fig:ideal_noise}(a) and (b) show the TVD and JSD, respectively, between the ground truth output of the Baseline circuit and the approximated noiseless output of \sol{}. The figure shows that both the output distance metrics have low values across all algorithms. Given the corresponding reduction in CNOT gate count, these results signify the usefulness of circuit approximations even in a fault-tolerant environment. As both metrics have similar trends, we use only the TVD going forward for brevity.

\begin{figure}[t]
    \centering
    \includegraphics[scale=0.47]{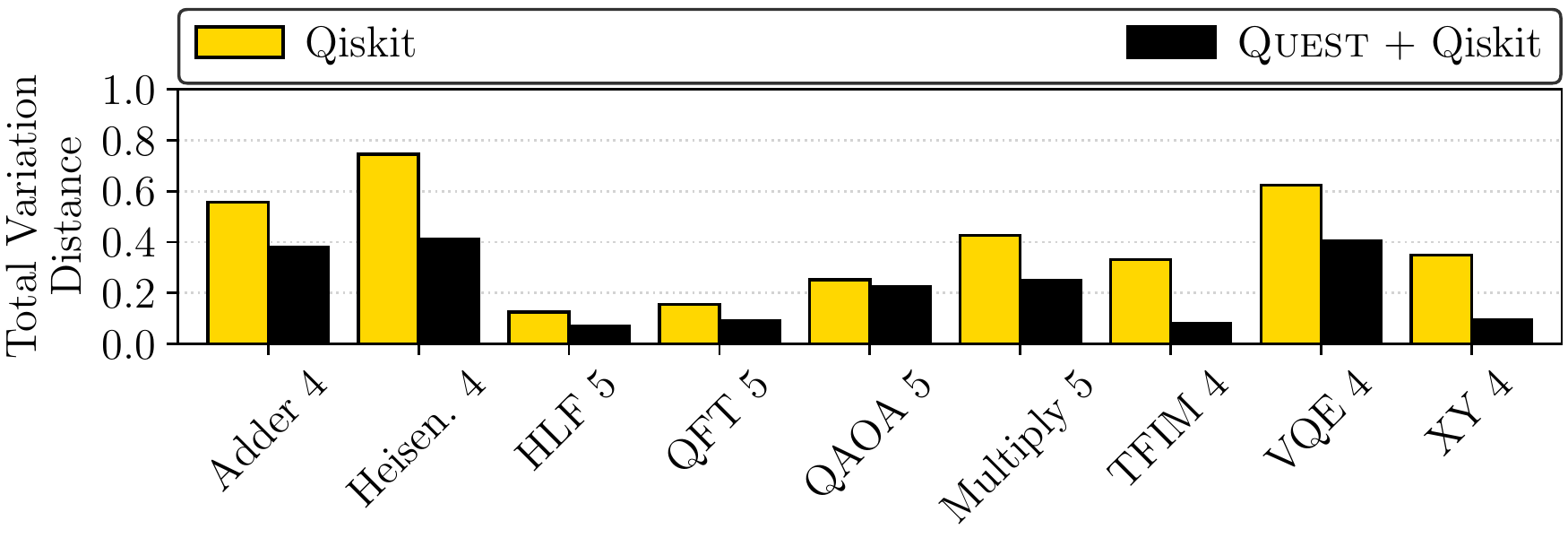}
    \caption{\sol{} + Qiskit reduces the TVD compared to Qiskit when the circuits are run on the IBMQ Manila computer.}
    \label{fig:real_machine}
  \end{figure}

\begin{figure*}[t]
    \centering
    \subfloat[][1\% Noise]{\includegraphics[scale=0.46]{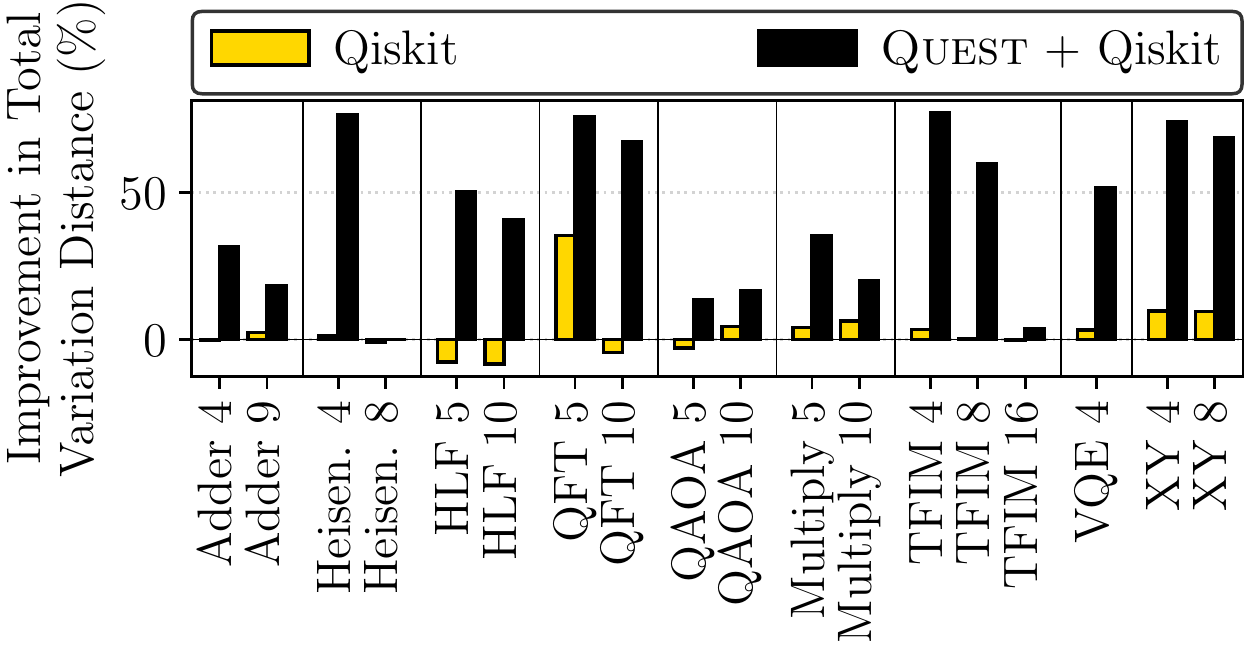}} \hfill
    \subfloat[][0.5\% Noise]{\includegraphics[scale=0.46]{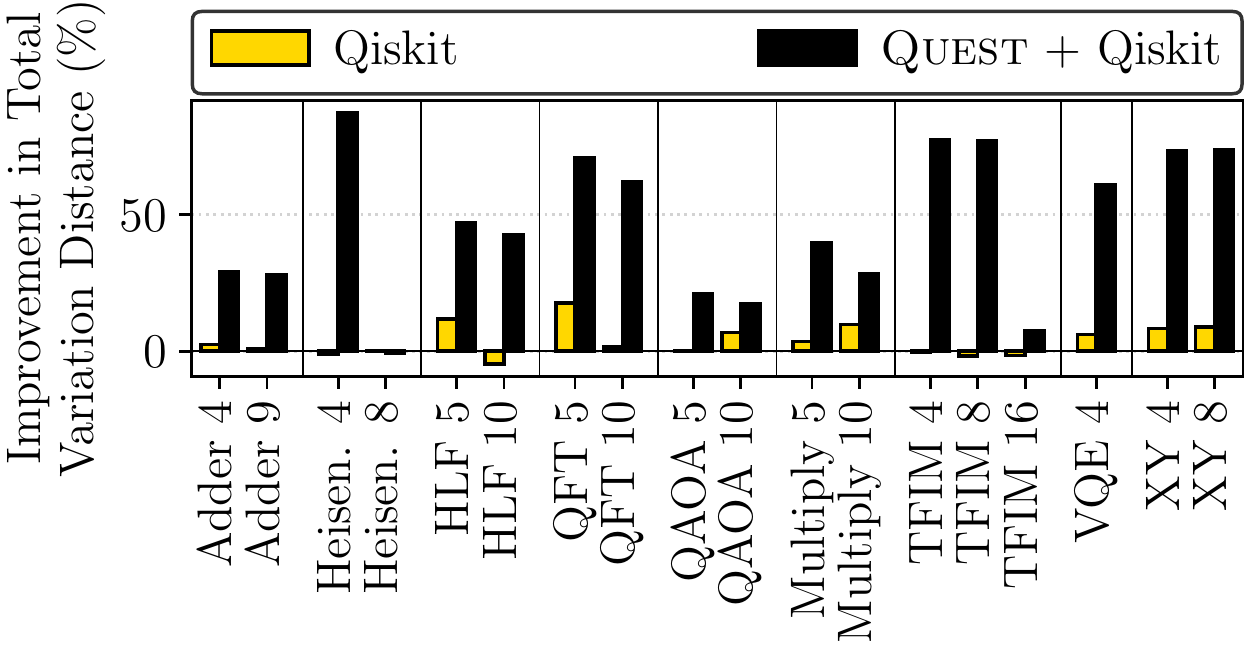}} \hfill
    \subfloat[][0.1\% Noise]{\includegraphics[scale=0.46]{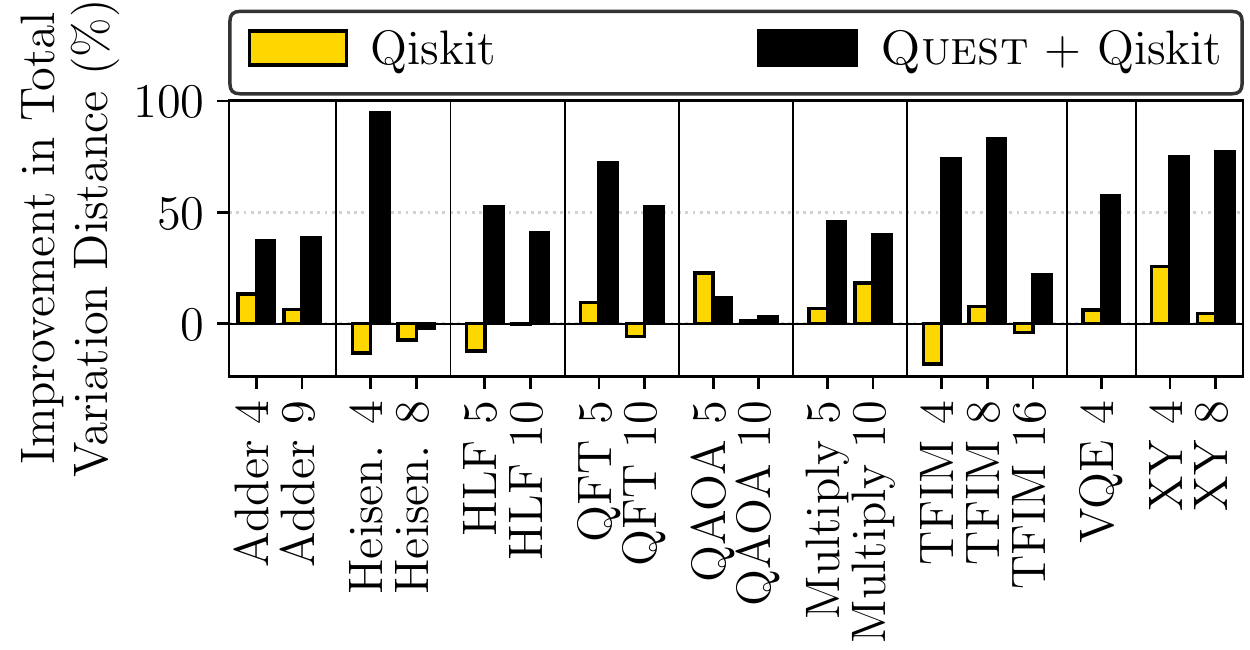}}
    \caption{Noisy simulations with different levels of noise indicate that \sol{} can reduce TVD even for future NISQ devices.}
    \label{fig:dec_noise}
  \end{figure*}

\noindent\textit{\newline \sol{} delivers a significant reduction in TVD on a real NISQ machine.} Next, we evaluate the performance of \sol{} on IBMQ Manila, one of the most recent and least error quantum computers available via IBMQ open access for algorithms which were possible to run. Fig.~\ref{fig:real_machine} shows the TVD from the ground truth when the algorithms are run with just the Qiskit optimizations vs. when they are run with \sol{} + Qiskit. While the raw TVD numbers are large for most algorithms due to the high noise of the current state-of-the-art quantum computers, \sol{} + Qiskit reduces the TVD by over 0.3 or 30\% points in some cases. For example, for the four-qubit TFIM circuit, the TVD drops from 0.35 to 0.08.

\begin{figure}[t]
    \centering
    \subfloat[][Execution Times]{\includegraphics[scale=0.48]{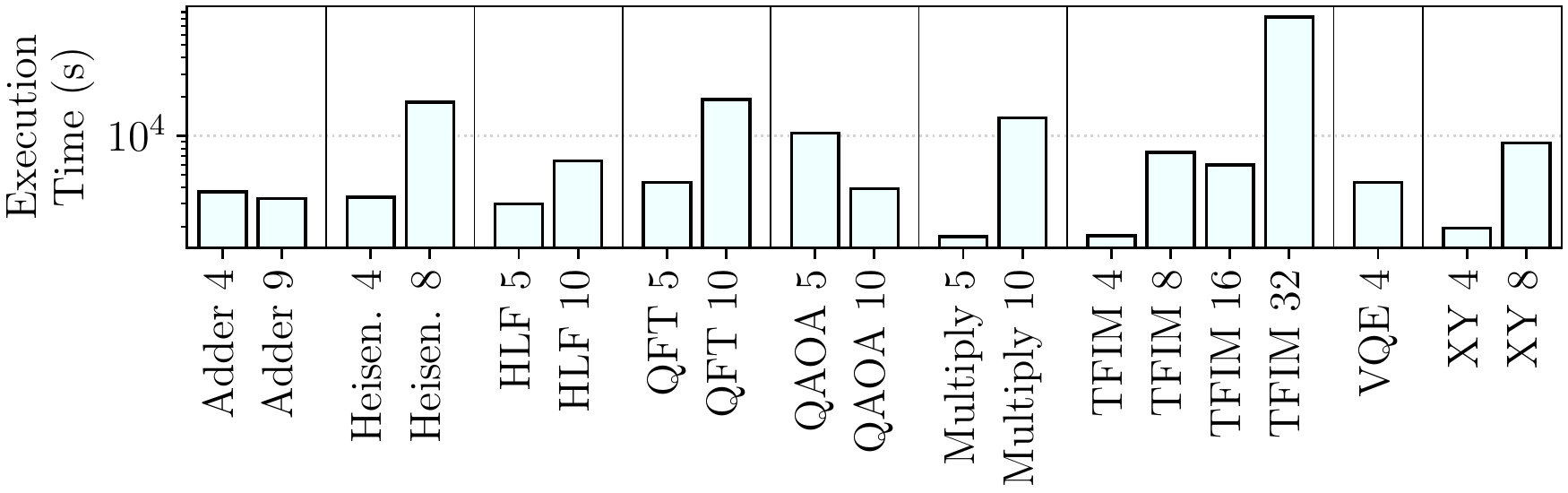}} \vspace{0mm}
    \subfloat[][Execution Time Division]{\includegraphics[scale=0.48]{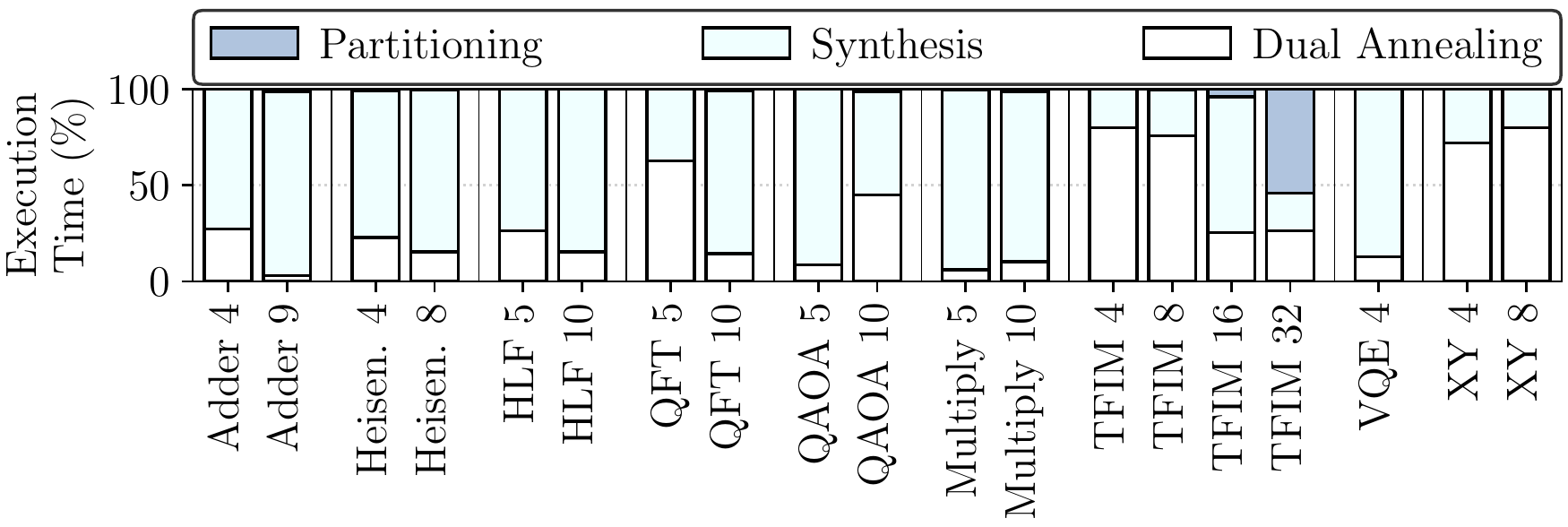}}
    \caption{The execution time overhead of \sol{} and its division among the different steps varies for different algorithms.}
    \label{fig:exec_time}
  \end{figure}

\noindent\textit{\newline \sol{} delivers a significant reduction in TVD even for larger circuits and as hardware noise is decreased in a quantum simulation.} We now present larger circuits run in a noisy simulation environment with noise levels of 1\%, 0.5\%, and 1\% in Fig.~\ref{fig:dec_noise}(a), (b), and (c), respectively. The figures show the percentage reduction in TVD compared to when the Baseline circuit is run with a noisy simulation for Qiskit and \sol{} + Qiskit. We see that across the board, \sol{} + Qiskit reduces the TVD even as the hardware noise is reduced. This demonstrates the usefulness of circuit approximations even when projected on to the future when the noise is reduced by 10$\times$ (0.1\% compared to the current average noise of over 1\%).


\noindent\textit{\sol{} incurs one-time cost for building approximate circuits to yield meaningful output quality for circuits with a large number of CNOT gates.} \sol{}’s approximate circuit building process consists of three steps: (1) partitioning, (2) synthesis, and (3) dual-annealing engine. Fig.~\ref{fig:exec_time}(a) and (b) show absolute time required for different circuit and relative contribution from each step. Overall, for most circuits \sol{} can be completed within a few hours. The only exception is TFIM 32 which takes almost a full day. Partitioning takes up most of the time due to TFIM circuit structure. Synthesis and dual-annealing engines are not major contributers.

While this one-time cost is non-negligible, it has potential for significant reduction (e.g., all blocks can be synthesized in parallel). But, we did not need to focus on reducing this overhead because of the feedback provided by the domain scientists and physicists who are actively working and improving the two major target applications. They assessed this cost is hidden in the code development and improvement cycle, and hence, wanted this effort to focus on obtaining meaningful output quality. For example, the magnetization curve for the Heisenberg application should match the ground truth curve. \sol{} achieved these algorithmic and science goals, as confirmed by our case study on TFIM and Heisenberg over their entire time evolution landscape (discussed below).

Also, we note that this overhead is not incurred each time the program needs to be compiled with Qiskit and executed. This is because \sol{} produces full approximate circuits as one-time output. This one-time output can be compiled with Qiskit (in the order of seconds) whenever needed for optimally mapping the approximate circuit on physical qubits.

\subsection{TFIM and Heisenberg: A Case Study}

\begin{figure}[t]
    \centering
    \subfloat[][TFIM 4]{\includegraphics[scale=0.43]{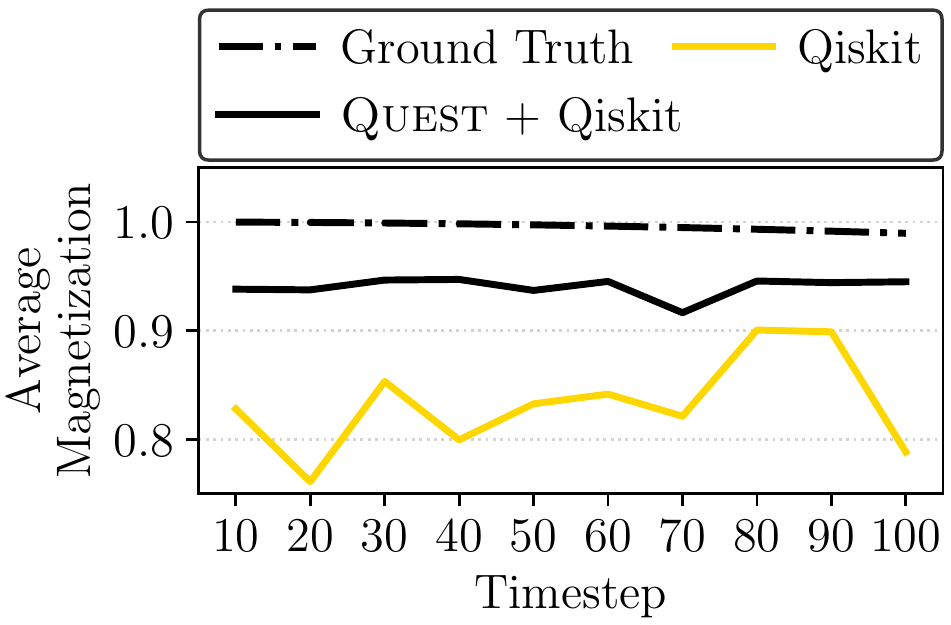}}\hfill
    \subfloat[][Heisenberg 4]{\includegraphics[scale=0.43]{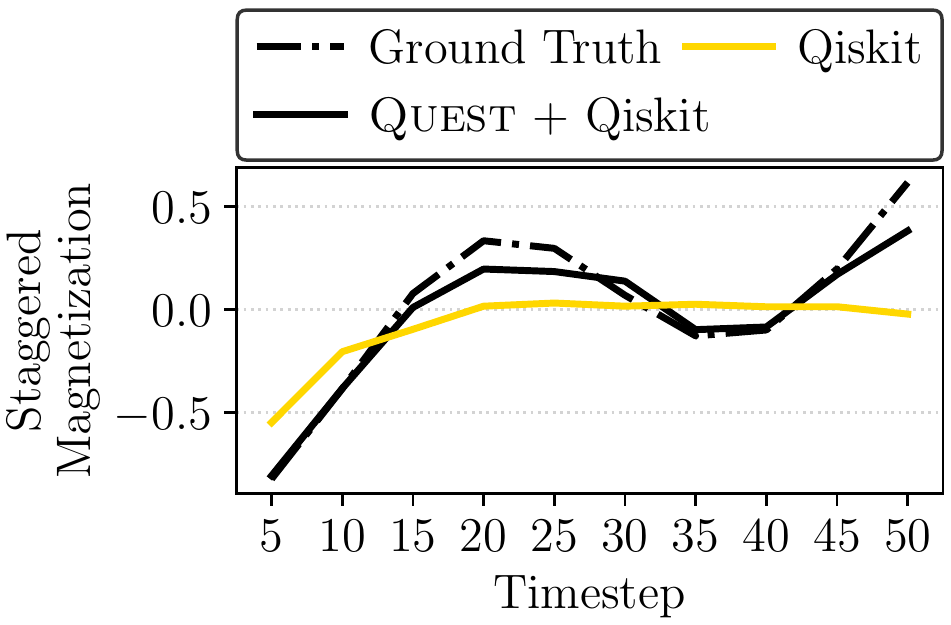}}
    \caption{\sol{} achieves closer to the ground truth output on the IBMQ Manila machine than the Baseline.}
    \label{fig:tfim_heis_real_mac}
  \end{figure}

\begin{figure}[t]
    \centering
    \subfloat[][TFIM 4]{\includegraphics[scale=0.43]{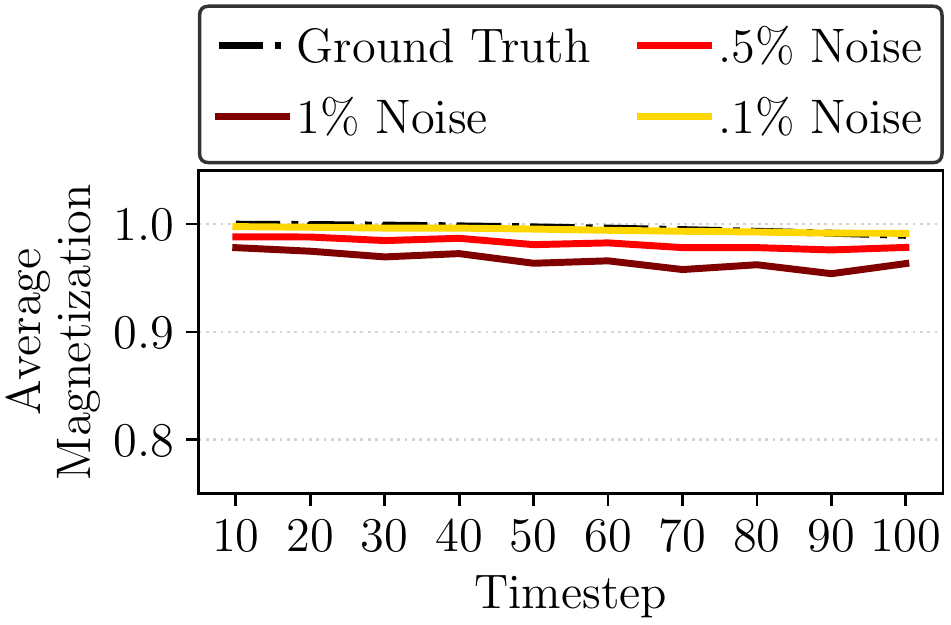}}\hfill
    \subfloat[][Heisenberg 4]{\includegraphics[scale=0.43]{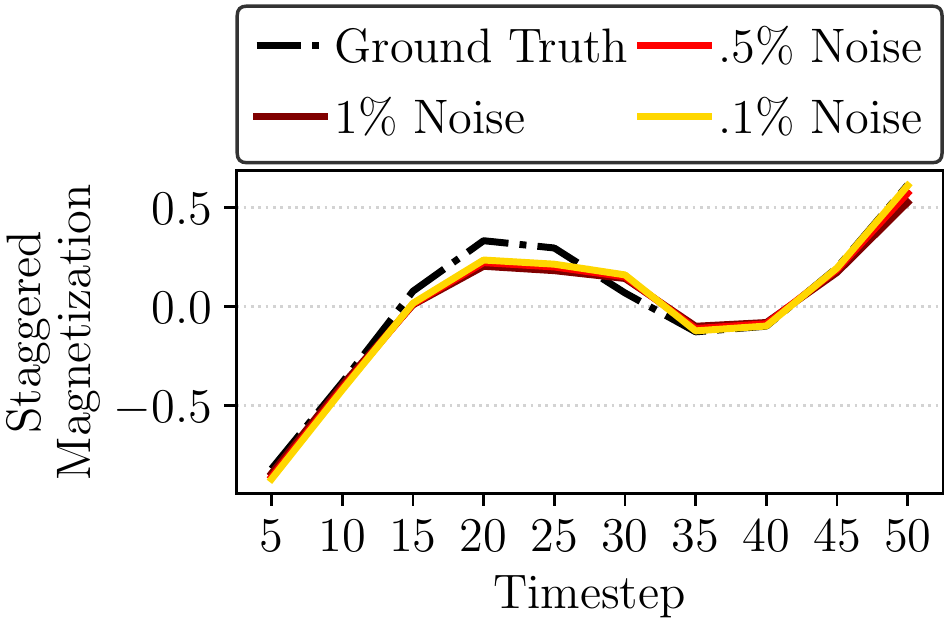}}\hfill
    \caption{For TFIM, \sol{} performs better as the hardware noise decreases. For Heisenberg, \sol{}'s output is close to ground truth even in a high noise environment of 1\%.}
    \label{fig:tfim_heis_step_noise}
  \end{figure}

\begin{figure}[t]
    \centering
    \subfloat[][TFIM 4]{\includegraphics[scale=0.3]{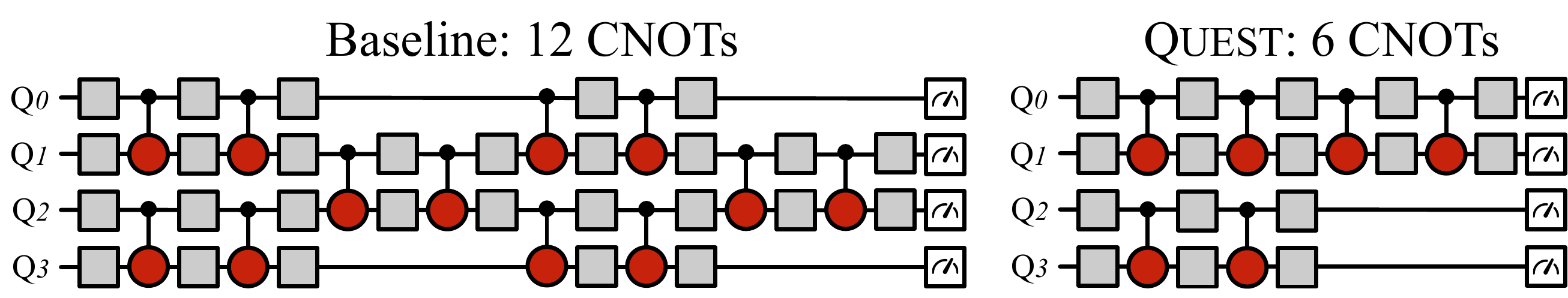}}\vspace{0mm}
    \subfloat[][Heisenberg 4]{\includegraphics[scale=0.3]{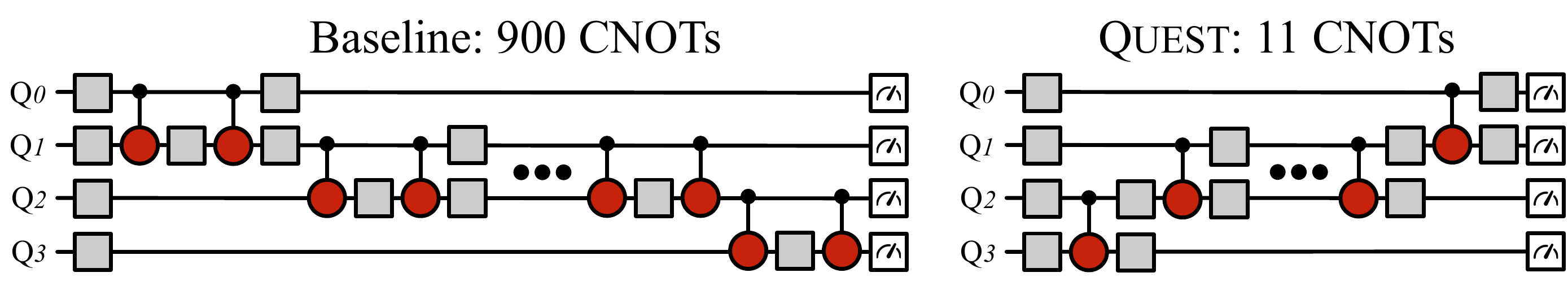}}
    \caption{Illustration of the reduction in the CNOT gate count for one of the TFIM and Heisenberg approximate circuits.}
    \label{fig:tfim_heis}
  \end{figure}

\noindent\textit{\sol{} + Qiskit is able to more closely track the ground truth output than just Qiskit optimizations due to the larger reduction in the number of CNOT gates.} Fig.~\ref{fig:tfim_heis_real_mac} shows the time evolutions of the four-spin TFIM and Heisenberg circuits with ground truth, and the Qiskit circuit, and \sol{} + Qiskit approximate circuits when run on the IBMQ Manila computer.  Each step in the time evolution is a different circuit that is separately run with \sol{}. For the TFIM algorithm, the \sol{} + Qiskit magnetization line is more stable and closer in magnitude to the ground truth than is the Qiskit magnetization line. For Heisenberg, \sol{} very closely tracks the ground truth magnetization, while Qiskit produces meaningless results.

In fact, Fig.~\ref{fig:tfim_heis_step_noise} plots the simulation results with noise levels 1\%, 0.5\%, and 0.1\%, and the figure shows that projected reduction in hardware errors can further reduce the output distance for TFIM and Heisenberg. This is due to the large reduction in the CNOT gate count of both the algorithms.

Fig.~\ref{fig:tfim_heis}(a) shows that circuit structure of the TFIM algorithm at the 100$^{th}$ timestep with Baseline and one of the approximations generated using \sol{}. Similarly, Fig.~\ref{fig:tfim_heis}(b) shows the circuit structure of the Heisenberg algorithm at the 50$^{th}$ timestep. The figures illustrate the large reduction in CNOT gate count. For example, for the Heisenberg algorithm, the gate count reduced from 900 CNOTs to just 11 CNOTs with approximate circuits. This reduction enables fewer operations errors and lower decoherence errors due to faster execution.

\begin{figure}[t]
    \centering
    \subfloat[][TFIM 4]{\includegraphics[scale=0.43]{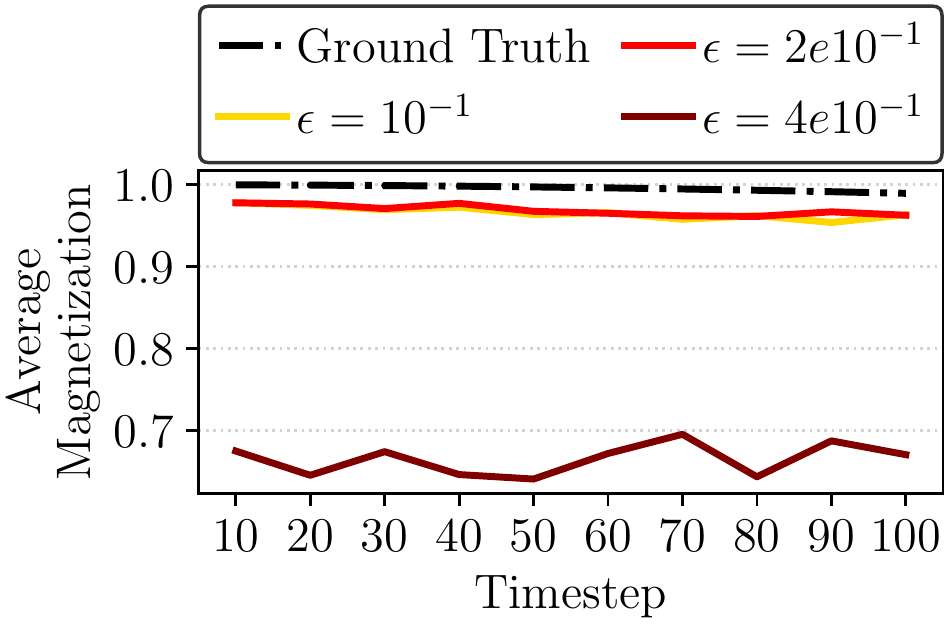}}\hfill
    \subfloat[][Heisenberg 4]{\includegraphics[scale=0.43]{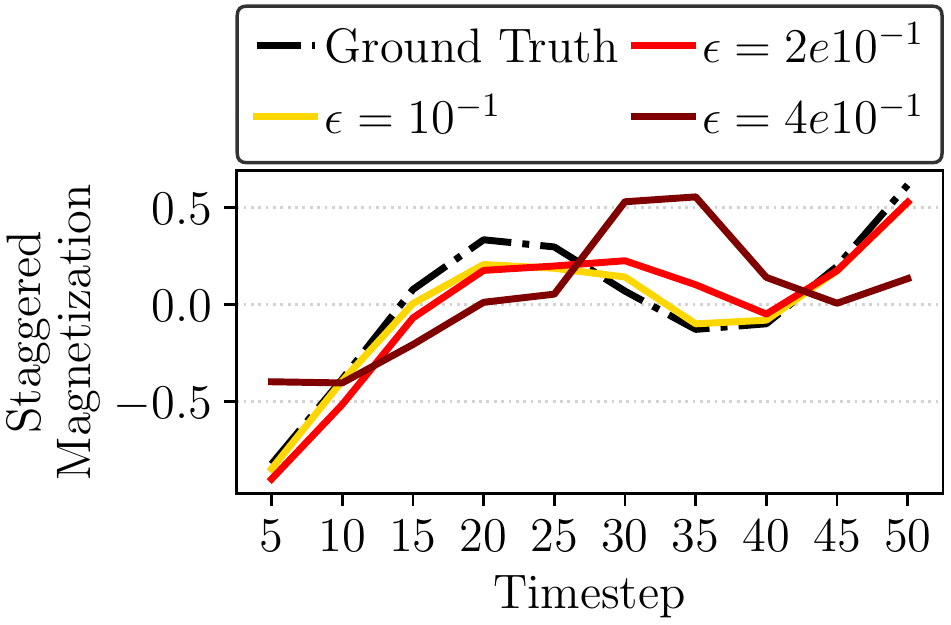}}\hfill
    \caption{Careful selection of process distance threshold for the dual annealing engine produces good approximations of the TFIM and Heisenberg circuits.}
    \label{fig:tfim_heis_step_bin}
  \end{figure}

\noindent\textit{\newline \sol{}'s design decisions of using process distance upper bound threshold and dissimilar approximations yields good results.} Fig.~\ref{fig:tfim_heis_step_bin}(a) and (b) shows the output of TFIM and Heisenberg algorithms for different process distance thresholds. Recall that if the threshold is set too high, the dual annealing engine is likely to select coarse approximations selecting more circuits with fewer CNOT gate count than selecting ones with dissimilar characteristics (approximations with low process distances among them). The figures shows that this can lead to a large error in the output distance for both algorithms. Thus, careful selection of the threshold is required to ensure beneficial results. However, it also does not have to be tuned exhaustively because as the figure shows, \sol{} performs well for a wide range of values. Therefore, we set the threshold to be proportional to the number of blocks in the circuit and it works well across all algorithms in practice.

\section{Related Work}
\label{sec:related_work}

\noindent\textbf{Quantum Circuit Compiling and Mapping.} There has been a large focus on attempting to leverage compiler-based passes and quantum computing rules to reduce CNOT and SWAP gate counts and perform noise- and layout- aware mapping of the qubits to the hardware~\cite{zhang2021time,murali2020software,murali2019noise,tannu2019not,shi2019optimized,li2019tackling,wille2019mapping,zulehner2019compiling,smith2019quantum}.

For example, in the circuit mapping space, previous efforts have exploited the diverse error characteristics of different qubits to map the same baseline circuit in different ways expecting the output distances to reduce~\cite{tannu2019ensemble,patel2020veritas}. But these works do not target reducing the CNOT count considerably by employing approximate synthesis to systematically generate dissimilar circuits and thus, have large output distances.

\noindent\newline\textbf{Quantum Circuit Synthesis.} \sol{} employs synthesis as one of the steps in its procedure to generate approximate circuits. Previous synthesis works have attempted to synthesize circuits with only specific gates (e.g., only CNOTs) or universal circuits as exactly as possible with as few CNOT gates as possible~\cite{shende2006synthesis,iten2016quantum,martinez2016compiling,iten2019introduction,kissinger2019cnot,nash2020quantum,davis2020towards,younis2021qfast,smith2021leap}. We use a modified version of the Leap synthesis tool for \sol{} as it performs better than previous approaches~\cite{smith2021leap}.

Despite the potential of approximations, not many procedures to generate resource efficient approximations using synthesis have existed before \sol{}. Madden et al.~\cite{madden2021best} describe generative procedures using synthesis, but these procedures are non-scalable and lack any apriori criteria for selecting approximations across different algorithms. Amy et al.~\cite{amy2013meet} describe a more scalable direct synthesis algorithm, but the approach leads to very long circuits, sometimes by orders of magnitude compared to other synthesis tools. In comparison, \sol{} defines a clear criterion for selecting dissimilar approximate circuits apriori, provides a theoretical proof to bound process distance, and generates circuits with few CNOTs.


  
\section{Conclusion}
\label{sec:conclusion}

In this work, we present \sol{}, a technique to reduce CNOT gate count of quantum circuits using approximate synthesis and distance-based approximation selection. We provided a theoretical derivation to bound the process distance of approximations as well as a dissimilarity criterion to select approximations in a manner that reduces the output distance. \sol{} achieves a CNOT gate reduction of 30-80\% across algorithms while maintaining a low output distance from the output of the original circuit. While \sol{}'s contributions are beneficial in the NISQ era for minimizing the impact of noise, they are also useful for fault-tolerant quantum computers.
\section{Acknowledgements}
\label{sec:ackowledgements}

This work was supported by the Office of Science, Office of Advanced Scientific Computing Research  Accelerated Research for Quantum Computing Program of the U.S. Department of Energy under Contract No. DE-AC02-05CH11231. This work was also supported by Northeastern University, NSF Award 1910601, and the Massachusetts Green High Performance Computing Center (MGHPCC) facility. This research used resources of the Oak Ridge Leadership Computing Facility, which is a DOE Office of Science User Facility supported under Contract No. DE-AC05-00OR22725. IBM Q was also used for this work. The views expressed are those of the authors and do not reflect the official policy or position of IBM or the IBM Q team.

\bibliographystyle{ACM-Reference-Format}
\bibliography{references}

\end{document}